\documentclass[aps,pre,reprint,superscriptaddress,longbibliography,floatfix,nofootinbib]{revtex4-2}

\usepackage{graphicx,amsmath,amssymb,mathtools, dcolumn,physics,color,wasysym, bm}
\usepackage[colorlinks= true, linkcolor=blue, citecolor=blue, urlcolor=blue, pdfusetitle]{hyperref}%

\graphicspath{{./figs/}{./}}

\newcommand{\revise}[1]{{\leavevmode\color{black}#1}}

\begin{document}

\title{\revise{Time-reversal symmetry breaking in the chemosensory array reveals mechanisms for dissipation-enhanced cooperative sensing}}

\author{David Hathcock}
\thanks{These two authors contributed equally to this work.} 
\affiliation{IBM T.~J.~Watson Research Center, Yorktown Heights, NY 10598}

\author{Qiwei Yu}
\thanks{These two authors contributed equally to this work.} 
\affiliation{IBM T.~J.~Watson Research Center, Yorktown Heights, NY 10598}
\affiliation{Lewis-Sigler Institute for Integrative Genomics, Princeton University, Princeton, NJ 08544}

\author{Yuhai Tu}
\affiliation{IBM T.~J.~Watson Research Center, Yorktown Heights, NY 10598}

\begin{abstract}

The \emph{Escherichia coli} chemoreceptors form an extensive array that achieves cooperative and adaptive sensing of extracellular signals.
The receptors control the activity of histidine kinase CheA, which drives a nonequilibrium phosphorylation-dephosphorylation reaction cycle for response regulator CheY. 
Cooperativity and dissipation are both important aspects of chemotaxis signaling, yet their consequences have only been studied separately.
Recent single-cell FRET measurements revealed that kinase activity of the array spontaneously switches between active and inactive states, with asymmetric switching times that signify time-reversal symmetry breaking in the underlying dynamics. 
Here, we present a nonequilibrium lattice model of the chemosensory array, which demonstrates that the observed asymmetric switching dynamics can only be explained by an interplay between the dissipative reactions within individual core units and the cooperative coupling between neighboring units. 
Microscopically, the switching time asymmetry originates from irreversible transition paths. 
The model shows that strong dissipation enables sensitive and rapid signaling response by relieving the speed-sensitivity trade-off, which %
can be tested by future single-cell experiments. 
Overall, our model provides a general framework for studying biological complexes composed of coupled subunits that are individually driven by dissipative cycles and the rich nonequilibrium physics within. 
\end{abstract}

\maketitle

Cellular sensing is resource intensive: signaling networks must be built to overcome the large fluctuations and small number statistics typically present in chemical signals. To deal with this noise, cells integrate repeated measurements, which are made using a large number of receptors on the cell surface \cite{berg1977physics,bialek2005physical}. The mechanisms for averaging and copying signals to intracellular carriers often involve burning fuel molecules, for example, phosphorylation cycles powered by ATP hydrolysis \cite{cao_free-energy_2015, goldbeter1981amplified,govern2014optimal,mehta2016landauer,ouldridge2017thermodynamics}. The relationships between cell resources (receptors, signaling proteins, energy) and sensing fidelity have been studied extensively \cite{ouldridge2017thermodynamics, mehta2016landauer, theWolde2016fundamental, govern2014optimal, lan_energyspeedaccuracy_2012, hathcock_nonequilibrium_2023,tjalma_trade-offs_2023}. In general, increasing the energy input raises the optimal speed and accuracy of cellular sensing. 
The receptors, however, do not sense independently: in \emph{Escherichia coli}, for example, CryoEM imaging \cite{briegel2009universal,liu_molecular_2012, Briegel2102Bacterial, cassidy2020structure} shows that chemoreceptors and the associated histidine kinase couple together in a hexagonal lattice and the downstream signal shows considerable cooperativity \cite{shimizu2010modular, Amin2010Chemoreceptors, tu2013quantitative, keegstra_phenotypic_2017}. Little is known about how the signaling response is impacted by the interplay between dissipative chemical cycles, driven out-of-equilibrium by ATP hydrolysis, and the collective behavior of the receptor lattice.

We address this question in the context of the \emph{E. coli} chemotaxis signaling pathway, where recent single-cell studies provided a new view into the nonequilibrium dynamics of chemoreceptor lattices \cite{keegstra_phenotypic_2017,keegstra_near-critical_2022}. In the absence of extracellular ligands and adaptation machinery, cells spontaneously switch between active and inactive states, which normally correspond respectively to no-ligand and large ligand concentration responses \cite{shimizu2010modular, Amin2010Chemoreceptors, keegstra_phenotypic_2017, tu2013quantitative}. This behavior is reminiscent of switching between metastable magnetized states in a finite-size Ising lattice except for one feature: the time to execute a switch from the active to inactive states is longer than the reverse process. Asymmetric switching breaks time-reversal symmetry, indicating that the collective dynamics of the chemoreceptor lattice are driven by a dissipative process. However, a quantitative description of the nonequilibrium dynamics at the receptor cluster level remains missing. 

We introduce a model of the chemoreceptor array that combines, for the first time, the nonequilibrium properties of individual receptor-kinase core units with unit-to-unit coupling in an extended lattice structure. Incorporating these two key features reveals collective nonequilibrium effects in dynamics and sensing response that are fundamentally different from equilibrium (e.g. Ising-like) models~\cite{Mello03, Lan2011Adapt, keegstra_near-critical_2022}. 

\begin{figure*}[t]
    \centering
    \includegraphics[width=0.9\linewidth]{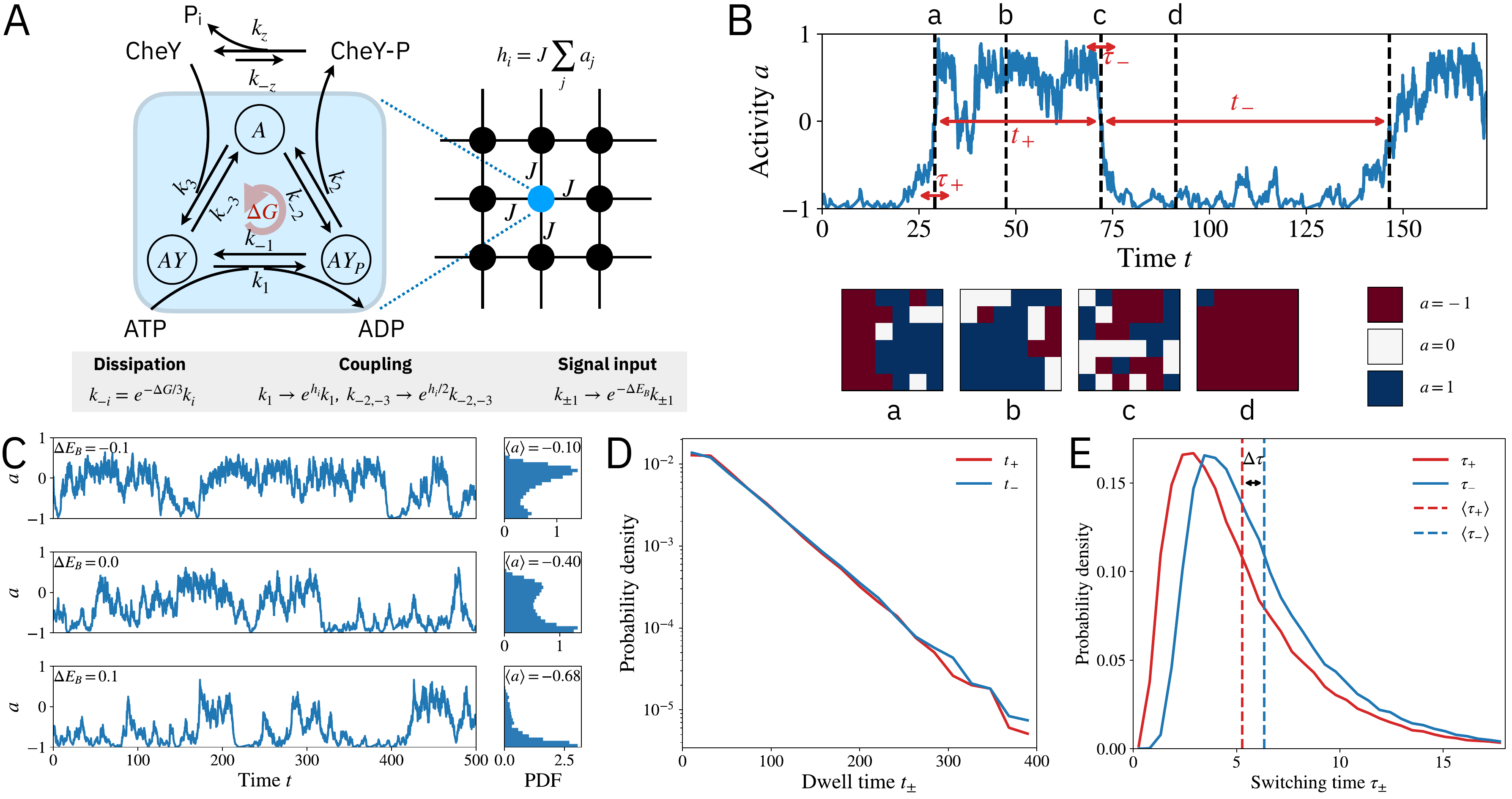}
    \caption{nonequilibrium lattice model of the chemoreceptor cluster.
    (A) Illustration of the nonequilibrium receptor cluster model. The kinases form a square lattice (right panel) with nearest neighbor coupling $J$. Each kinase is in one of the three states (blue box) with transition rates indicated on the arrows. \revise{The rates are controlled by three effects described in the grey box--see main text for details.} 
    (B) Typical trajectory of the total activity (upper) and the corresponding snapshots (lower).
    (C) Activity trajectories (left) and histograms (right) for different barrier heights $\Delta E_B$.
    (D) Dwell time distributions for both active and inactive states are exponential.
    (E) Switching time distributions are asymmetric: switching to the inactive state takes longer. Vertical dashed lines show the means of each distribution.
    Parameters for (C)--(E): $k_3=0.5$, $J=1.18$, $\epsilon=e^{-\Delta G/3}=0$. Lattice size $N=6\times 6$.
    }
    \label{fig1:model}
\end{figure*}

The nonequilibrium lattice displays a rich array of dynamics not possible in equilibrium: we show that dissipative driving is a fundamental requirement for the asymmetric switching observed in experiments. This result, along with the magnitude of the measured dwell times between switching events, indicates that the chemosensory array operates near a highly nonequilibrium critical point. 
Microscopically, asymmetric switching arises because the lattice follows different paths through state-space depending on the direction of the switch: inactive to active or vice versa, another tell-tale consequence of time-reversal symmetry breaking in the underlying system.   
Beyond switching dynamics, our study shows that operating out of equilibrium enhances both the amplitude and the speed of the sensing response, thus allowing the chemosensory array to respond swiftly and sharply to changes in the ligand concentration. This high-fidelity sensing behavior sharply contrasts the response of equilibrium lattices: fueled by energy dissipation, the chemosensory array can ease, or even reverse, speed-sensitivity trade-offs that are fundamental to equilibrium mechanisms~\cite{sartori_free_2015,fei_design_2018, lan_energyspeedaccuracy_2012}.
More broadly, our model establishes a framework for describing a variety of biophysical systems comprising coupled subsystems, each driven out of equilibrium.

\section{nonequilibrium lattice model of the chemoreceptor cluster}
Our nonequilibrium lattice model is inspired by the chemoreceptor arrays found in \emph{E. coli}, whose core functional unit comprises two trimers of chemoreceptor dimers, one histidine kinase CheA dimer, and two coupling proteins CheW~\cite{parkinson_signaling_2015}. Binding of external signals (ligands) to the chemoreceptors controls the kinase activity of CheA, i.e., its ability to phosphorylate the response regulator CheY. The phosphorylation-dephosphorylation (PdP) cycle is completed by CheZ catalyzing the dephosphorylation of CheY-P.
The core units form an extended hexagonal lattice on the plasma membrane \cite{briegel2009universal,liu_molecular_2012, Briegel2102Bacterial, cassidy2020structure}. 
The kinase activity of the core unit depends on both the occupancy of ligands and the methylation level of the receptors. Recent experimental evidence~\cite{Amin2010Chemoreceptors,Levit02,Vaknin07} favors a nonequilibrium allosteric model~\cite{hathcock_nonequilibrium_2023}, which explicitly includes the PdP cycle driven by ATP hydrolysis. 
nonequilibrium driving is necessary to explain the disproportionate shifts in ligand binding and kinase response due to receptor methylation~\cite{Amin2010Chemoreceptors,Levit02,Vaknin07,hathcock_nonequilibrium_2023}. 
Here, we generalize this model to an extended lattice by incorporating the PdP cycle of individual kinase with the cooperative interactions between neighboring core units.

We use a minimal three-state model to describe the kinase activity of core unit-$i$: $a_i=0$, $-1$, and $1$ represent the unbound kinase $\mathrm{A_i}$, the kinase bound to unphosphorylated CheY ($\mathrm{A_iY}$), and the kinase bound to CheY-P ($\mathrm{A_iY_P}$), respectively\footnote{In reality there are more sub-steps in the phosphorylation cycle, including CheA autophosphorylation and phosphotransfer to CheY. 
Here, we opt for the simplest nonequilibrium model with a coarse-grained three-state cycle.}.  As illustrated in Fig.~\ref{fig1:model}A, for each of the $N$ core units in the lattice, the kinase activity dynamics can be described by the following reaction cycle,
\revise{
\begin{equation}
\begin{split}
    \text{Binding: }& \text{A}_i + \text{Y} \xrightleftharpoons[k_{-3} ]{k_3} \text{A}_i\text{Y}, \\
    \text{Phosphorylation: }& \text{A}_i\text{Y} +\text{ATP} \xrightleftharpoons[k_{-1} ]{k_1 } \text{A}_i \text{Y}_\text{P} +\text{ADP},\\
    \text{Unbinding: }& \text{A}_i \text{Y}_\text{P}\xrightleftharpoons[k_{-2} ]{k_2} \text{A}_i +\text{Y}_\text{P},
\end{split}
\end{equation}
}where A and Y, and Y$_\text{P}$ represent CheA, CheY and CheY-P respectively, and $i=1,2,\dots N$ labels the core units. 
\revise{The transition rates $k_{\pm i}$ ($i=1,2,3$) are controlled by three effects: energy dissipation, coupling (cooperativity) between neighboring signaling units, and allosteric signaling due to ligand binding and receptor methylation (grey box in Fig.~\ref{fig1:model}A), which we describe in the following. 

Energy dissipation due to ATP hydrolysis drives each core unit out of equilibrium. Completing the PdP cycle consumes one ATP molecule with free energy dissipation 
\begin{equation}
    \Delta G_\mathrm{ATP}= \Delta G + \Delta G_z= \ln \frac{k_1 k_2 k_3}{k_{-1}k_{-2}k_{-3}}+\ln \frac{k_z}{k_{-z}},
\end{equation}
with $\Delta G$ and $\Delta G_z$ being the contributions from the kinase cycle of the core unit and the dephosphorylation by CheZ, respectively. 
For studying the dependence of lattice dynamics on dissipation, we assume the reverse transition rates scale identically: $k_{-i} = k_i e^{-\Delta G/3} \equiv k_i \epsilon$. 

The core units in the chemosensory array sense cooperatively, which is captured by nearest neighbor coupling with strength $J$ (Fig.~\ref{fig1:model}A). We use a square lattice for simplicity but the switching behavior is robust to lattice structure and size (see SI, section II). 
The effective field due to nearest neighbor kinase $h_i = J \sum_j a_j$ scales the forward phosphorylation rate for kinase $i$: $k_1\to k_1 e^{h_i}$. To maintain thermodynamic consistency in the PdP cycle, $k_{-2}$ and $k_{-3}$ are also scaled by $e^{h_i/2}$. 
We set $k_2=1$ to fix the unit of time.
CheY-P is dephosphorylated in solution at rate $k_z$ with the reverse rate $k_{-z}$.
}

\revise{The kinase activity of CheA is controlled allosterically by ligand occupancy and methylation state of the receptors. A recently proposed nonequilibrium allosteric model~\cite{hathcock_nonequilibrium_2023} suggests that ligand binding and receptor methylation affect kinase activity in different ways}: binding to the receptors acts as an ON/OFF switch for the kinase, while methylation shifts the energy barrier for the phosphorylation reaction in the ON state. Together these effects can be captured by an effective barrier shift $\Delta E_B$ that scales the forward and reverse phosphorylation rates: $k_{\pm1} = k_{\pm 1}^0 \exp(-\Delta E_B)$, where $k_{\pm 1}^0$ are the rates corresponding to equal occupancy of the active and inactive states. %
If the relation between receptor conformation and occupancy is described by the Monod-Wyman-Changeux (MWC) model~\cite{MWC1965, hathcock_nonequilibrium_2023}, the effective barrier shift is
\begin{equation}\label{Eq:mwc}
    \Delta E_B(m, [L]) = n \qty[\alpha  (m-m_0) +  \ln \left(\frac{1+[L]/K_i}{1+[L]/K_a} \right)].
\end{equation}
The first term corresponds to the direct barrier shift due to methylation $m$, with slope $\alpha$ and intercept $m_0$.
The second term describes the effective barrier shift due to the fraction of bound receptors: $[L]$ is the ligand concentration and $K_i$ ($K_a$) is the dissociation constant for the inactive (active) receptor.
Both contributions are proportional to $n$, the effective receptor cluster size felt by the kinase within a core unit.
In analogy to the Ising model, $\Delta E_B$ plays the role of an external field that modulates the activity of the entire lattice. 
\revise{In this study, we isolate the lattice response properties by directly tuning $\Delta E_B$, which can describe the response to both $[L]$ and $m$. Measurements of core-unit response~\cite{li2014selective, pinas2016source} will be useful for determining the phenomenological MWC parameters in Eq.~(\ref{Eq:mwc}) or a more precise microscopic model of the core unit in future work (see Discussion).}

\begin{figure*}[t]
    \centering
    \includegraphics[width=0.65\linewidth]{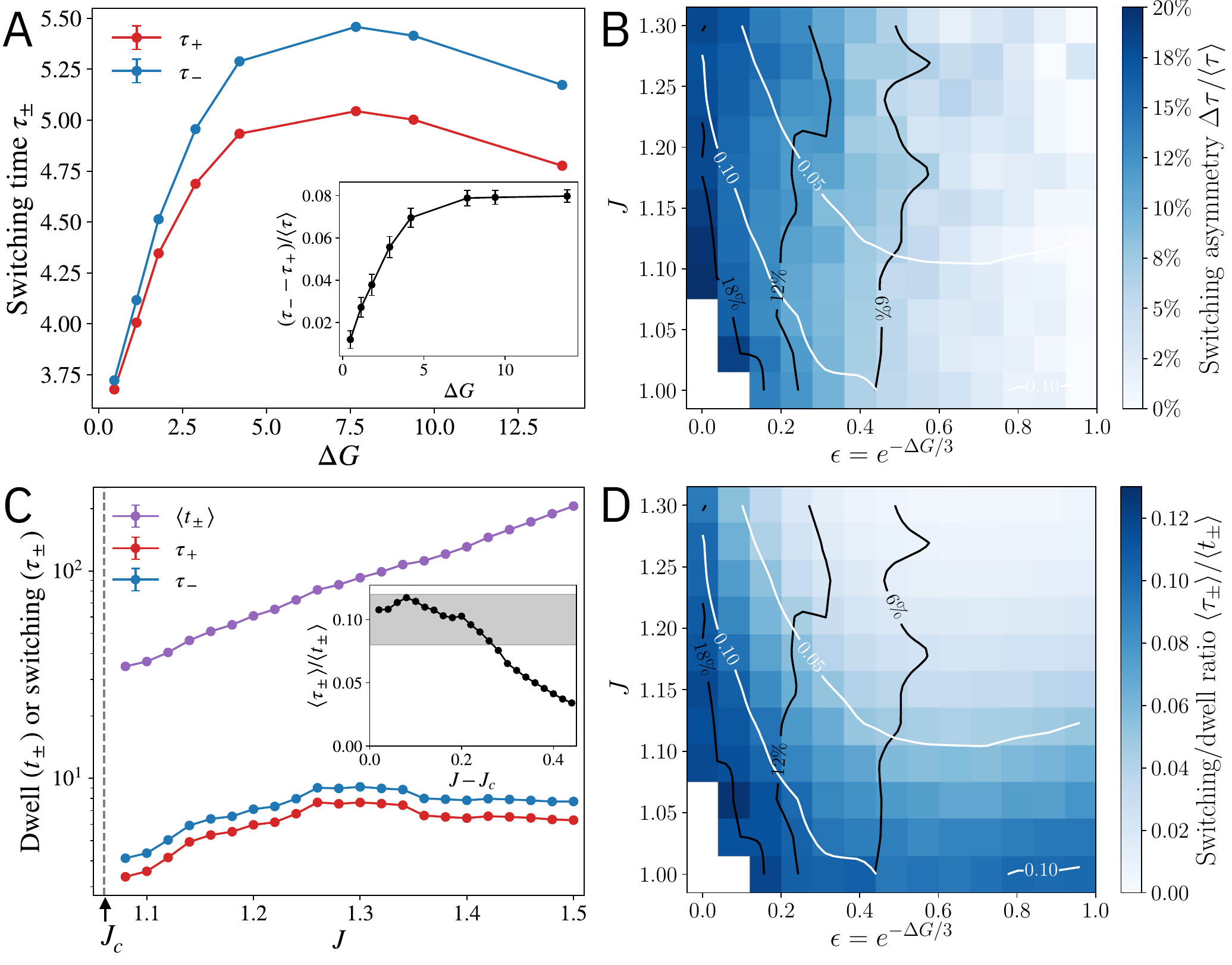}
    \caption{Time statistics of two-state switching. 
         (A) Switching times and switching asymmetry (inset)
         as a function of dissipation $\Delta G$.
         (B) Switching time asymmetry as a function of $J$ and $\epsilon=e^{-\Delta G/3}$. 
         (C) Switching and dwell times and the switching/dwell time ratio (inset) as a function of lattice coupling coupling $J$. The shaded region in the inset shows the range of experimentally measured switching/dwell time ratio~\cite{keegstra_near-critical_2022}.  
         (D) Switching/dwell time ratio as a function of $J$ and $\epsilon=e^{-\Delta G/3}$. 
         For (B) and (D), the white bins lie below the critical coupling $J_c(\epsilon)$ with no two-state switching. Contours show constant asymmetry (black) and switching/dwell ratio (white). Experimentally measured switching statistics require large dissipation ($\epsilon \approx 0$) and proximity to the critical point ($J\gtrsim J_c$).
         Parameters: (A)  $J=1.1$, $k_3=1.0$; (C) $\epsilon=0.01$, $k_3=0.5$; (B, D) $k_3=0.5$.
         \revise{Lattice size $N=6\times 6$. The qualitative behavior of asymmetric switching is independent of lattice size (see SI section II C-D).}
    }
    \label{fig2:switching time}
\end{figure*}

The dynamics of the lattice are simulated using the Gillespie algorithm~\cite{gillespie_exact_1977}.
We focus on the kinase states and do not simulate the dephosphorylation step explicitly; instead, the intracellular concentrations of CheY, CheY-P, ATP, and ADP  are held constant and absorbed into pseudo-first-order reaction rates. 
With a sufficiently strong coupling ($J>J_c$), the system undergoes two-state switching similar to the measured activity switching in \emph{E. coli}~\cite{keegstra_near-critical_2022}. 
As shown in Fig.~\ref{fig1:model}B, the two-state switching behavior can be readily detected in the time trace of the order parameter (lattice activity) $\expval{a} = N^{-1} \sum_i a_i$, along with snapshots of the lattice at various times.
Following the experimental literature~\cite{keegstra_near-critical_2022}, we define dwell times $t_{\pm}$ to be the time spent in the active and inactive states respectively, and the switching times $\tau_{\pm}$ as the duration of transition between these two states (Fig.~\ref{fig1:model}B). 
The average activity can be controlled by modulating the effective barrier $\Delta E_B$ (via ligand concentration and/or receptor methylation): lowering the energy barrier enhances activity (Fig.~\ref{fig1:model}C). 
Consistent with experiments, the dwell times are exponentially distributed (Fig.~\ref{fig1:model}D), while the switching time distributions are peaked, right skewed, and asymmetric (Fig.~\ref{fig1:model}E).

\section{Chemoreceptor array operates near a highly nonequilibrium critical point}

The nonequilibrium lattice model lets us identify the key ingredients that lead to the asymmetric two-state switching observed in \emph{E. coli} chemoreceptor arrays~\cite{keegstra_near-critical_2022}. 
In mutants containing only serine-binding receptors (Tsr) or aspartate receptors (Tar), and for cells without activity bias (i.e. equal dwell times in the active and inactive states), the measurements found a switching-to-dwell-time ratio of $0.08-0.12$ and a switching time asymmetry of $25-35\%$~\cite{keegstra_near-critical_2022}. As illustrated below, the large asymmetry implies that the system operates far from equilibrium, while the switching-to-dwell-time ratio evidences proximity to criticality.

\textbf{The observed switching time asymmetry requires strong dissipation.}
The switching time distribution is determined by the ensemble of transition paths: trajectories that go from one state (e.g. inactive) to the other (active) without returning to the source~\cite{e_transition-path_2010}.
For equilibrium systems, time-reversal symmetry establishes a one-to-one correspondence between forward and backward trajectories. Thus, the transition path ensembles are equivalent, resulting in identical switching time distributions. 
This symmetry has been shown explicitly in specific cases~\cite{berezhkovskii_forwardbackward_2019}. 
Therefore, the disparity between forward and backward switching time distributions in the \emph{E. coli} measurements~\cite{keegstra_near-critical_2022} and in our model (Fig.~\ref{fig1:model}) is a clear signature of the underlying dissipative driving breaking time-reversal symmetry. 
Here, the driving is provided by ATP hydrolysis through $\Delta G$: indeed, reducing dissipation reduces the switching time asymmetry (Fig.~\ref{fig2:switching time}A). 

Varying both dissipation and coupling strength (Fig.~\ref{fig2:switching time}B) reveals that the switching time asymmetry depends only weakly on coupling strength $J$, while strong dissipation ($\epsilon = \exp(-\Delta G/3)\ll 1$) is always required. 
In the large dissipation limit, the model explains a $20\%$ asymmetry. The remaining asymmetry observed in the experiments may be accounted for by considering the dephosphorylation reaction, neglected in our simulations. For example, the difference between the timescales for dephosphorylation and phosphorylation with a fully active receptor cluster is approximately $0.5$s~\cite{Sourjik02a}, accounting for about $10\%$ additional asymmetry~\cite{keegstra_near-critical_2022}. Furthermore, the asymmetry may be underestimated because the coarse-grained three-state reaction cycle used here neglects intermediate chemical reactions. Coarse-graining is known to lower the dissipation rate in chemical networks~\cite{yu2021inverse,yu_state-space_2022} and hence may also decrease switching asymmetry.

\textbf{The observed switching-to-dwell-time ratio indicates proximity to criticality.} 
Two-state switching only emerges when the coupling is above a critical strength ($J>J_c$). 
This critical coupling depends both on the dissipation level and the competition of timescales determined by kinetic rates. 
As we show in the mean-field limit (see SI, section I), the critical coupling $J_c$ in general varies non-monotonically with the dissipation.

In equilibrium barrier crossing problems, the dwell time grows exponentially with the energy barrier~\cite{hanggi_reaction-rate_1990} while the transition time grows much slower~\cite{hummer_transition_2004,chung_experimental_2009}. Here we find similar relations: as the coupling strength increases away from the critical point, the dwell times grow exponentially 
while the switching times saturate for large $J$ (Fig.~\ref{fig2:switching time}C). The switching-to-dwell-time ratio shrinks exponentially away from the critical point (inset). 
Therefore, the measured ratio between switching and dwell times provides an indirect measure of how strongly the chemoreceptors are coupled. 

Varying both dissipation and coupling reveals that the switching-to-dwell-time ratio depends primarily on the distance of the coupling strength from its critical value $J_c(\epsilon)$ and decays exponentially as $J$ grows (Fig.~\ref{fig2:switching time}D). 
To observe spontaneous switching on an experimental timescale, the cells must therefore operate in reasonable proximity to the critical point: otherwise, dwell times become exponentially long compared to the switching time and can not explain the observed switching-to-dwell-time ratios, $\expval{\tau_\pm}/\expval{t_\pm} \approx 0.10$.

\revise{The qualitative features of the asymmetric two-state switching described in this section are preserved for larger lattice models, across the range of the size of native chemoreceptor arrays (see SI section II C-D). In particular, we find that there is always a moderate range of coupling strengths near the critical point for which two-state switching occurs with a switching-to-dwell-time ratio comparable to the experimentally measured range (0.08-0.12). While the switching asymmetry decreases with system size, the magnitude and scaling with $N$ depend on kinetic features (e.g. the number of intermediate enzyme states and the kinetic rates) and connectivity (lattice model versus mean-field limit ). 
It is therefore likely that generalizations of our model that incorporate more details of the kinetic network and biological structure could fully quantitatively explain the experimentally observed switching statistics even in large lattices (see Discussion). 
}

\section{Microscopic origin of asymmetric switching}
\begin{figure}[t]
    \centering
    \includegraphics[width=0.85\linewidth]{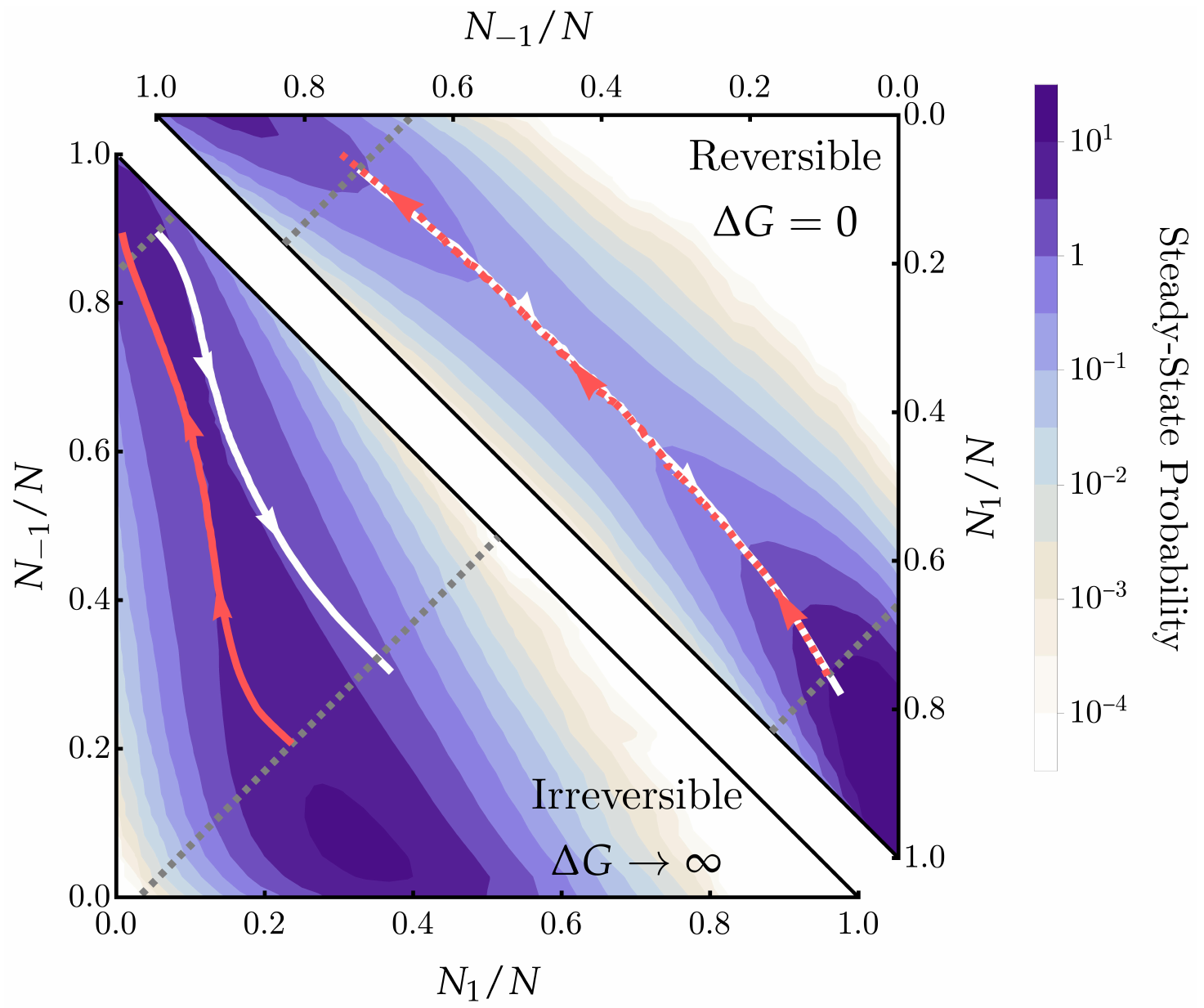}
    \caption{
        Irreversible switching paths underlie the switching time asymmetry. 
        Mean forward (white) and backward (red) switching trajectories from lattice simulations projected onto the $(N_1, N_{-1})$--plane.
        The lower triangle shows the extreme irreversible limit ($\Delta G \to \infty$) with asymmetric switching paths, and the upper triangle shows the reversible limit ($\Delta G=0$) with symmetric paths. 
        The background contours show the steady-state probability distribution. 
        Active and inactive regions of state-space are defined using fixed activity thresholds with constant $\langle a \rangle = (N_1-N_{-1})/N$ (gray dashed lines). Trajectories the mean-field limit, for which the state-space is exactly the ($N_1$, $N_{-1}$)--plane, are qualitatively identical (see SI). }
    \label{fig3:path}
\end{figure}

Our model provides insight into the microscopic origin of the switching time asymmetry: depending on the direction of switching, transitions take distinct paths through the lattice space. 
This can be illustrated by projecting the average transition paths onto a plane spanned by the fraction of core units in the AY\textsubscript{P} ($N_1/N$) and AY ($N_{-1}/N$) states, respectively. 
In the irreversible limit (Fig.~\ref{fig3:path}, lower triangle), trajectories toward the inactive state (red) tend to take paths through the lower left region of state space (where the number of $a=0$ core-units, $N_0 = N-N_{-1}-N_1$ is large), whereas trajectories toward the active state (white) run closer to the diagonal. This behavior is intuitive because, in the underlying three-state cycle (Fig.~\ref{fig1:model}A), the core unit moves through the 0-state on the way to the $-1$ inactive state, while the transition from $-1$ to $1$ is direct. The difference in switching trajectories is discernible from snapshots of the lattice simulations (Fig.~\ref{fig1:model}B): when switching to inactive a considerable fraction of the kinase occupies the $0$-state compared to the forward switches to the active state.

Keegstra et al. found switching-time distributions were well described by gamma distributions, $\text{Gamma}(\alpha, \beta)$~\cite{keegstra_near-critical_2022}, which is the distribution of times to execute an $\alpha$-step process with rate $\beta$ for each step. Thus, $\alpha$ can be used as a proxy for the number of independent timescales underlying the switching process. The larger fit $\alpha$ for switching to the inactive state versus switching to the active state (2.45 versus 1.72 for Tsr and 2.74 versus 1.87 for Tar~\cite{keegstra_near-critical_2022}) reflects the presence of more mixed time-scales in the backward switch, consistent with our model.

In the absence of dissipative driving, the state-space trajectories during switching are identical (Fig.~\ref{fig3:path}, upper triangle), a consequence of time-reversal symmetry. 
In this case, the microscopic cycles are symmetric (Fig.~\ref{fig1:model}A):
the core units are equally likely to transition clockwise or counter-clockwise regardless of the direction of switching.

\begin{figure*}[t]
    \centering
    \includegraphics[width=\linewidth]{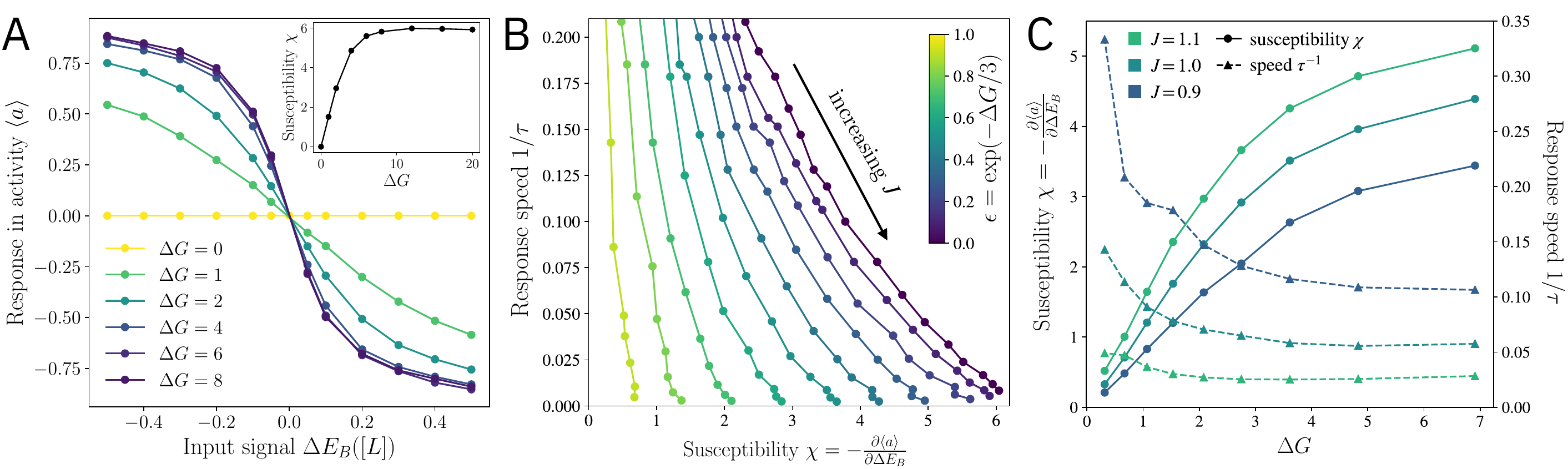}
    \caption{
        The activity response functions of the nonequilibrium chemosensory array to external signals $[L]$, transmitted via modulating the effective phosphorylation energy barrier $\Delta E_B([L])$. 
        (A) The change in activity $\expval{a}$ due to an external signal $\Delta E_B([L])$ for various dissipation levels $\Delta G$. The inset shows the susceptibility  $\chi = - \eval{\pdv{\expval{a}}{\Delta E_B}}_{\Delta E_B=0}$.
        (B) Linear response speed $\tau^{-1}$ versus susceptibility $\chi$. 
        Each line represents a fixed dissipation level $\Delta G$ (see inset legend) with different couplings $J$.
        Large dissipation improves the speed-amplitude trade-off. 
        (C) At fixed coupling $J$ and in the linear regime ($\Delta E_B \ll 1$), increasing dissipation $\Delta G$ enhances susceptibility (solid lines) at the cost of response speed (dashed lines). 
        For (B--C), susceptibility is computed by $\chi=-\frac{\Delta\expval{a}}{\Delta E_B}$ with a small applied field $\Delta E_B=-0.1$;
        $\tau$ is obtained by fitting the exponential relaxation of the activity $\expval{a}(t)=\expval{a}_\infty+\Delta \expval{a} e^{-t/\tau}$.
        We use $k_3=2.5$ in all panels.  %
    }
    \label{fig4:response}
\end{figure*}

\begin{figure*}[t]
    \centering
    \includegraphics[width=0.7\linewidth]{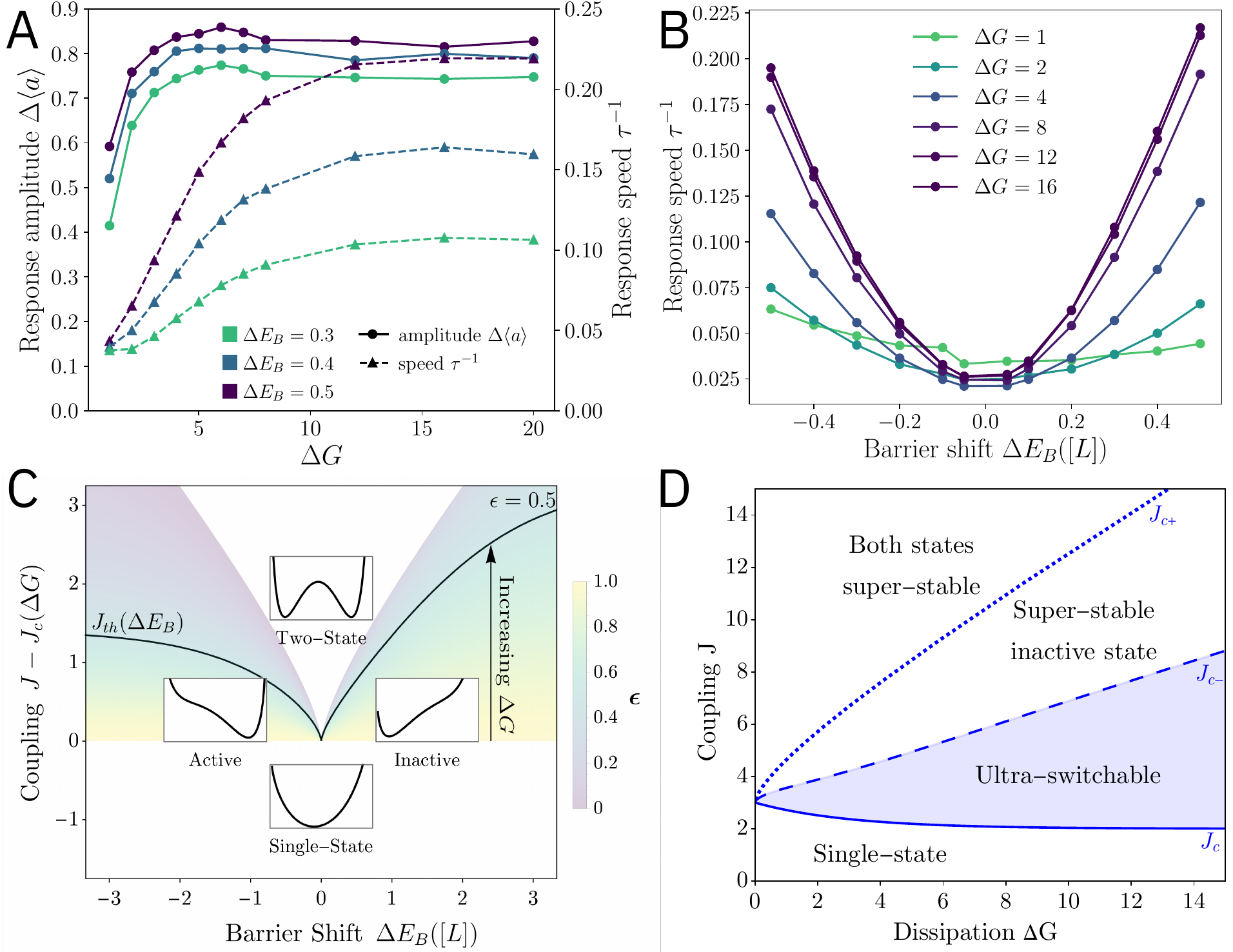}
    \caption{
       Nonlinear response and phase diagrams. 
       (A) Response amplitude $\Delta \expval{a}$ (solid lines) and response speed $\tau^{-1}$ (dashed lines) as functions of dissipation $\Delta G$ at large external signals $\Delta E_B([L])$. 
       (B) The response speed as a function of the barrier shift $\Delta E_B([L])$ for different dissipation levels.
       (C) Heat map of the phase boundaries in the $(\Delta E_B, J)$-plane for different dissipation levels. The black line indicates the boundary $J_{th}(\Delta E_B)$ for $\epsilon=0.5$. 
       The insets illustrate the effective free energy landscape $F=-\ln P$: above the phase boundary the system exhibits two-state switching (two minima), while below it has a single state (one minimum) with activity depending on the barrier shift $\Delta E_B$. 
       For $\epsilon>0$ ($\Delta G$ finite) the boundary has a finite asymptote $J_{th}(\Delta E_B) \to J_{c\pm}$ as $\Delta E_B \to \pm \infty$. 
       (D) Phase boundaries (blue) in the $(\epsilon, J)$-plane for two-state switching $J_c$ and super-stable states $J_{c\pm}$ ($k_3 \gg 1$ limit). Operating out of equilibrium ($\Delta G \gg 1$) gives the largest range of coupling where the system is \emph{ultra-switchable}: with bistability that can be eliminated by a barrier shift in either direction. Tuning the balance of timescales (e.g. via $k_3$) shifts the boundaries, but the \emph{ultra-switchable} region still expands with increasing dissipation (see SI, section I).
    }
    \label{fig5:phase diagram}
\end{figure*}

\section{Dissipation enhances the sensing response properties of the chemosensory array}

The spontaneous asymmetric switching considered above was observed in cells without extracellular ligands and adaptation machinery~\cite{keegstra_near-critical_2022}. How does the nonequilibrium mechanism underlying the asymmetry affect the sensing properties of the receptor cluster? 
Recent studies \cite{hathcock_nonequilibrium_2023, Amin2010Chemoreceptors,Levit02,Vaknin07} provide strong evidence that the receptor complex acts as an enzyme exerting kinetic control: both ligand concentration and receptor methylation affect the kinase activity by modulating the effective phosphorylation energy barrier (Eq.~\ref{Eq:mwc}).
Thus, the signal input is a shift in the effective barrier $\Delta E_B$ induced by changes in the extracellular ligand concentration $[L]$.
The output is the intracellular CheY-P concentration, which is proportional to the fraction of active kinase. %
Here we focus on the response on a much shorter timescale than adaptation feedback. 

In our model, $\Delta E_B$ controls the overall lattice activity $\expval{a}$\footnote{The response functions for other order parameters, e.g. $\expval{N_1/N}$ or $\expval{N_{-1}/N}$ have nearly identical behaviors to all results in this section (see SI, section II).} with a sigmoidal response curve (Fig.~\ref{fig4:response}A) reminiscent of \emph{E. coli's} activity response to varying ligand concentration \cite{shimizu2010modular, Amin2010Chemoreceptors, keegstra_phenotypic_2017}. 
Consistent with previous work~\cite{hathcock_nonequilibrium_2023}, the response amplitude increases with dissipation and vanishes in the equilibrium limit, where the steady state is unaffected by the kinetic barrier shift. 
The zero-field susceptibility $\chi = - \eval{\pdv{\expval{a}}{\Delta E_B}}_{\Delta E_B=0}$ also increases monotonically with dissipation (Fig.~\ref{fig4:response}A, inset). The sharp response in the large dissipation limit is reminiscent of a nonequilibrium ultrasensitivity mechanism proposed for the \textit{E. coli} flagellar motor switch~\cite{tu2008nonequilibrium}. 
In the following sections, we demonstrate that dissipation enhances response both in the linear (small $\Delta E_B$) and nonlinear (large $\Delta E_B$) regimes.

\textbf{Linear regime: dissipation improves the speed-sensitivity trade-off}. 
Small signals ($\Delta E_B\ll 1$) induce changes in activity proportional to $\Delta E_B$, with susceptibility $\chi$ defined above. 
The sensitivity to changes in ligand concentration can be quantified by an effective Hill coefficient $H=2 \mathrm{d}\log \expval{\tilde{a}}/\mathrm{d}\log [L]$, with $\expval{\tilde{a}}=(a_\mathrm{max}-\expval{a})/(a_\mathrm{max}-a_\mathrm{min})$ being the normalized activity.
Using Eq.~(\ref{Eq:mwc}), the maximum sensitivity is (Methods),
\begin{equation}
    H_\text{max} = 4A^{-1} n \chi,
\end{equation} 
where $A = a_\mathrm{max}-a_\mathrm{min}$ 
is the total response range.
The response speed is defined as $\tau^{-1}$, with the relaxation time $\tau$ obtained by fitting the activity to an exponential relaxation process.

Tuning the coupling $J$ leads to a trade-off between speed and sensitivity (Fig.~\ref{fig4:response}B). At strong coupling, a large response amplitude arises from flipping the entire system altogether (e.g. from all-active to all-inactive). However, this comes at the cost of slow speed since the dwell time in each state increases exponentially with coupling (Fig.~\ref{fig2:switching time}B). 
Increasing dissipation $\Delta G$ improves the trade-off, pushing the curves to the upper right and allowing for higher sensitivity at a fixed response speed (and vice versa). Moreover, dissipation also decreases the slope of the trade-off curves so that increasing sensitivity has a smaller cost in speed. 

Varying dissipation at fixed coupling leads to another speed-sensitivity trade-off (Fig.~\ref{fig4:response}C): increasing dissipation amplifies the response amplitude (solid lines) at the cost of slower response speed (dashed lines). Therefore, improving the trade-off in Fig.~\ref{fig4:response}B requires simultaneously varying $J$ and $\Delta G$: for example, enhancing sensitivity at a fixed response speed requires increasing dissipation $\Delta G$ while reducing coupling $J$ to compensate for the loss in speed. Beyond dissipation and coupling, tuning the relative timescales in the phosphorylation cycle (e.g. by modulating $k_3$) can also improve the speed-sensitivity trade-off (see SI, section II).

\textbf{Nonlinear regime: dissipation enables sensitive and rapid response.} 
In the nonlinear regime, with sufficiently large $\Delta E_B$, the barrier in the effective energy landscape disappears. This leads to an even more favorable situation in which dissipation not only increases the response amplitude $\Delta \expval{a}$ but also speed $\tau^{-1}$ (Fig.~\ref{fig5:phase diagram}A). %
Indeed, sweeping across a wide range of dissipation and barrier shifts (Fig.~\ref{fig5:phase diagram}B) reveals that dissipation enables a significant speed-up in response in the nonlinear regime. %

The speed enhancement in the nonlinear regime can be understood analytically by using mean-field theory. As shown in the ($J$, $\Delta E_B$)-plane (Fig.~\ref{fig5:phase diagram}C),
for a given $\Delta G$, the system exhibits two stable states when the coupling strength is larger than some threshold $J_{th}$: $J>J_{th}(\Delta E_B,\Delta G)$ (black line). The threshold coupling $J_{th}$ is determined by %
the onset of a saddle-node bifurcation that eliminates the energy barrier between the two states (see SI, section I).
For $J>J_{th}$, a large signal can cause the system to cross the phase boundary and transition to a fully active or inactive state rapidly since the energy barrier is eliminated. %
As $\Delta G$ increases (from yellow to blue), $J_{th}$ increases, thereby reducing the signal ($\Delta E_B$) required to cross the phase boundary and trigger a rapid response.

It is, however, not always possible to cross the phase boundary.
The phase boundaries for finite $\Delta G$ have finite asymptotes $J_{c\pm}\equiv J_{th}(\Delta E_B \to \pm\infty)$ (see SI, section I). 
Above $J_{c\pm}$, the kinase active or inactive states become super stable: they cannot be destabilized by any external ligand signal. %
In this case, the response, even to very strong signals, is slow because it is limited by the dwell times of the two stable states. 

The two limiting thresholds $J_{c+}$ and $J_{c-}(<J_{c+})$ along with $J_c \equiv J_{th}(\Delta E_B \to0)$, divide the parameter space into four regions %
(Fig.~\ref{fig5:phase diagram}D). 
The \emph{``ultra-switchable''} region $J\in(J_c, J_{c-})$ (shaded) is the most desirable since it offers rapid cooperative response to both activating or inhibiting signals.
Conversely, systems in the region $J\in(J_{c-}, J_{c+})$ respond slowly to activating signals, while beyond $J_{c+}$, the response is slow in both directions.
Increasing dissipation opens up the ultra-switchable region, enabling rapid response that is robust to perturbations in the coupling strength. %

\section{Discussion}
\textbf{Energy dissipation eases trade-offs between biological functions.} Our results show that the dynamics and response properties of nonequilibrium lattices are fundamentally different from those of equilibrium lattices: dissipation enables time-reversal symmetry breaking in fluctuation-driven spontaneous switching and enhances sensing fidelity. In particular, the nonequilibrium lattice model explains the observed asymmetric spontaneous activity switching in \emph{E. coli}~\cite{keegstra_near-critical_2022,keegstra_phenotypic_2017}. The dissipation underlying these dynamics improves the speed-sensitivity trade-off in chemotaxis signaling. Our work therefore supports an emerging general principle: operating out of equilibrium as a mechanism for relieving trade-offs between important biological functions.
On its own, energy dissipation is responsible for enabling ultrasensitivity in many biological systems~\cite{tu2008nonequilibrium}. Optimizing sensitivity, however, often comes with a trade-off for other functional goals such as noise suppression~\cite{sartori_free_2015,fei_design_2018} and collective motion~\cite{yu_flocking_2022}. By breaking the fluctuation-dissipation theorem~\cite{martin_comparison_2001,sartori_free_2015}, dissipation has been shown to ease these trade-offs, thereby enabling high sensitivity and low noise in biochemical oscillators~\cite{fei_design_2018}, as well as high sensitivity and speed in flocking~\cite{yu_flocking_2022}.

The speed-sensitivity trade-off predicted by our model can be experimentally tested by correlating single-cell kinase response measurements (amplitude and speed) with two-state switching measurements. For cells exhibiting unbiased switching in the absence of ligands~\cite{keegstra_near-critical_2022}, the ratio between switching and dwell times is a proxy for the coupling strength $J$. We expect this coupling is negatively correlated with response speed but positively correlated with sensitivity as shown in Fig.~\ref{fig4:response}B.

Another non-trivial prediction from our model is the speed-up in response for larger dosages of chemoattractants, a counterintuitive phenomena analogous to the Mpemba effect~\cite{lasanta_when_2017}. This could similarly be tested by measuring time-dependent kinase response to different dosages of chemoattractants or repellents. It will be interesting to see whether the observed speed-ups follow the roughly quadratic shape shown in Fig.~\ref{fig5:phase diagram}B. For large systems the quadratic coefficient diverges as the coupling approaches the critical point; therefore, this measurement may provide an alternative characterization of whether \emph{E. coli} operates near criticality.

Further, it may be interesting to search for superstable states by comparing response or switching dynamics in the presence of attractants and repellents. This measurement would be most interesting when combined with partial ATP depletion (e.g. using NaCN~\cite{koler2024signal}), since the superstable states are more prominent in the low dissipation regime (Fig.~\ref{fig5:phase diagram}D). These quantitative measurements would provide key new insights into how dissipation array structure affects chemotaxis response.

\textbf{Incorporating microscopic structure for high-fidelity models of the chemosensory array.}
\revise{Looking forward, our model serves as a starting point for developing high-fidelity models of the chemosensory system, both in \emph{E. coli} and similar bacteria, based on the detailed structural information from CryoEM imaging~\cite{briegel2009universal, liu_molecular_2012, Briegel2102Bacterial, cassidy2020structure}. Incorporating these details should not generally impact the qualitative behavior of the model: the underlying nonequilibrium cycles still break time-reversal symmetry and this dissipation will enhance signaling performance. However, such models will reveal the quantitative impact of biological structure on signaling function: in particular, the resulting cooperativity, response speed, adaptation accuracy, and the influence of dissipation on each of these key properties. 

To this end, a number of recent experiments provide quantitative data that will constrain parameters (such as the coupling strengths between different components of the array) in structure-based models. Structural details can be incorporated at multiple levels: both within the core unit (which we have taken as the nodes in our lattice model) and in the coupling between core units. For the former, the response of mixed receptor (Tar and Tsr) core complexes to both serine and aspartate have been measured for all possible compositions of trimers of dimers~\cite{li2014selective}. These measurements should clarify how different combinations of receptors impact the effective barrier shift [Eq.~(\ref{Eq:mwc})] that propagates ligand and methylation signals to the kinase.

The kinase-baseplate can assemble in multiple ways: most commonly observed is the canonical p6-symmetric hexagonal lattice structure, but various bacteria (including \emph{E. coli}) also exhibit arrays with p2 symmetry~\cite{muok2020atypical, burt2021alternative}. While we have shown here that lattice structure does not qualitatively impact the dynamics of the system, it may play an important role in determining the level of cooperativity in signaling response. In particular, in p2 arrays core units combine with a single ring structure via CheA-CheW interactions, while p6 arrays have an additional six-membered CheW ring, whose presence enhances cooperativity and sensitivity~\cite{pinas2022hexameric}.
The CheW hexamer therefore introduces a secondary coupling between core units, which could easily be added to our model to study its signaling implications. Because the CheW ring interacts with a specific receptor in the trimer of dimers, this secondary coupling could also be incorporated alongside the detailed core-unit model described above.

The role of CheW in chemosensory array dynamics is not fully understood. It has been established, for example, that lowering CheW expression levels leads to disorder in the cluster structure, which in turn lowers the downstream kinase response~\cite{khursigara2011lateral}. There is evidence that similar changes in structure can be actively induced by the signaling state: long duration of sustained ligand stimuli causes the arrays to locally fall apart, due to disruption of the interface II coupling between CheW and CheA~\cite{frank_prolonged_2013, pinas2016source}. This effect could be introduced into our model by adding a weak state dependence to the coupling strength. Feedback in the coupling may have interesting interactions with the dissipative driving from the phosphorylation-dephosphorylation cycle, which will be an exciting topic for future study. Whether these array dynamics play a functional role in signal processing remains an intriguing open question. Finally, modification of coupling constants impacts the activity bias in our model, which may provide an indirect mechanism for sensing of non-conventional ligands that do not bind to receptors. 

Besides the physical structure of the array, there is also an opportunity to incorporate increasingly detailed models of the underlying chemical reaction cycles. Specifically, the phosphorylation pathway involves a more intricate set of reactions than the three-state cycle considered here, including auto-phosphorylation of CheA and phosphotransfer to CheY, which occur on different domains of CheA \cite{mello_dual_2018, muok_regulation_2020}. These details introduce additional sources of irreversibility into the dynamics, which may explain the larger switching asymmetry observed in experiments compared to our simplified model. Incorporating the full reaction cycle into a lattice model will also shed new light on how specific components of the phosphorylation pathway are modified by ligand binding and methylation.
}

\textbf{Nonequilibrium lattices for modeling biological function and critical phenomena.}
Beyond chemotaxis, our nonequilibrium lattice model provides a novel framework for studying the dynamics and response properties of biological complexes composed of interacting subunits that are individually driven out of equilibrium by a dissipative cycle. In particular, the connectivity structure and kinetic network can be adapted for modeling a variety of biophysical systems. %
Potential applications include a broader range of signal transduction networks involving receptor clustering, such as EGF-receptors~\cite{schreiber_biological_1983} or G-protein coupled receptors \cite{sanchez2023ligand}, 
critical cell membrane dynamics~\cite{honerkamp-smith_experimental_2012}, actomyosin cortex~\cite{tan_scale-dependent_2021}, and large protein complexes such as the cytoplasmic ring in bacterial flagellar motor~\cite{HBerg2003}, the cooperative KaiC hexamer~\cite{Han2023KaiC} in circadian clock of Cyanobacteria and other ATPase~\cite{fang_empirical_2022}.

By introducing a minimal model for an extended lattice with coupled dissipative units, this work provides a platform for addressing crucial theoretical questions. 
For example, how does the subunit dissipative cycle affect the emergent dynamics and signaling capabilities of the entire lattice? 
Would it be possible to employ a nonequilibrium renormalization group approach~\cite{hohenberg_theory_1977,tu_renormalization_2023} to characterize the critical points and identify universality classes? 
The answer to these questions may generate important insights into nonequilibrium phase transitions driven by microscopic dissipative cycles.

\section{Additional information}
\textbf{Code availability.} The code for simulating the nonequilibrium lattice model using Gillespie algorithm is available in the GitHub repository: \href{https://github.com/qiweiyuu/NoneqLatticeChemotaxis}{https://github.com/qiweiyuu/NoneqLatticeChemotaxis}.

\textbf{Acknowledgements.}~This work is supported in part by National Institutes of Health grant R35GM131734 (to Y.~T.). 
Q.~Y.~acknowledges the IBM Exploratory Science Councils for a summer internship during which part of the work was done.

\textbf{Competing interests.}~The authors declare no competing interests.

\textbf{Author contributions.}~D.~H., Q.~Y., and Y.~T. designed research, performed research, and wrote the paper. D.~H. and Q.~Y. contributed equally.

\bibliography{ref_chemotaxis}

\begin{thebibliography}{60}%
\makeatletter
\providecommand \@ifxundefined [1]{%
 \@ifx{#1\undefined}
}%
\providecommand \@ifnum [1]{%
 \ifnum #1\expandafter \@firstoftwo
 \else \expandafter \@secondoftwo
 \fi
}%
\providecommand \@ifx [1]{%
 \ifx #1\expandafter \@firstoftwo
 \else \expandafter \@secondoftwo
 \fi
}%
\providecommand \natexlab [1]{#1}%
\providecommand \enquote  [1]{``#1''}%
\providecommand \bibnamefont  [1]{#1}%
\providecommand \bibfnamefont [1]{#1}%
\providecommand \citenamefont [1]{#1}%
\providecommand \href@noop [0]{\@secondoftwo}%
\providecommand \href [0]{\begingroup \@sanitize@url \@href}%
\providecommand \@href[1]{\@@startlink{#1}\@@href}%
\providecommand \@@href[1]{\endgroup#1\@@endlink}%
\providecommand \@sanitize@url [0]{\catcode `\\12\catcode `\$12\catcode
  `\&12\catcode `\#12\catcode `\^12\catcode `\_12\catcode `\%12\relax}%
\providecommand \@@startlink[1]{}%
\providecommand \@@endlink[0]{}%
\providecommand \url  [0]{\begingroup\@sanitize@url \@url }%
\providecommand \@url [1]{\endgroup\@href {#1}{\urlprefix }}%
\providecommand \urlprefix  [0]{URL }%
\providecommand \Eprint [0]{\href }%
\providecommand \doibase [0]{https://doi.org/}%
\providecommand \selectlanguage [0]{\@gobble}%
\providecommand \bibinfo  [0]{\@secondoftwo}%
\providecommand \bibfield  [0]{\@secondoftwo}%
\providecommand \translation [1]{[#1]}%
\providecommand \BibitemOpen [0]{}%
\providecommand \bibitemStop [0]{}%
\providecommand \bibitemNoStop [0]{.\EOS\space}%
\providecommand \EOS [0]{\spacefactor3000\relax}%
\providecommand \BibitemShut  [1]{\csname bibitem#1\endcsname}%
\let\auto@bib@innerbib\@empty
\bibitem [{\citenamefont {Berg}\ and\ \citenamefont
  {Purcell}(1977)}]{berg1977physics}%
  \BibitemOpen
  \bibfield  {author} {\bibinfo {author} {\bibfnamefont {H.~C.}\ \bibnamefont
  {Berg}}\ and\ \bibinfo {author} {\bibfnamefont {E.~M.}\ \bibnamefont
  {Purcell}},\ }\bibfield  {title} {\bibinfo {title} {Physics of
  chemoreception},\ }\href {https://doi.org/10.1016/S0006-3495(77)85544-6}
  {\bibfield  {journal} {\bibinfo  {journal} {Biophysical journal}\ }\textbf
  {\bibinfo {volume} {20}},\ \bibinfo {pages} {193} (\bibinfo {year}
  {1977})}\BibitemShut {NoStop}%
\bibitem [{\citenamefont {Bialek}\ and\ \citenamefont
  {Setayeshgar}(2005)}]{bialek2005physical}%
  \BibitemOpen
  \bibfield  {author} {\bibinfo {author} {\bibfnamefont {W.}~\bibnamefont
  {Bialek}}\ and\ \bibinfo {author} {\bibfnamefont {S.}~\bibnamefont
  {Setayeshgar}},\ }\bibfield  {title} {\bibinfo {title} {Physical limits to
  biochemical signaling},\ }\href {https://doi.org/10.1073/pnas.0504321102}
  {\bibfield  {journal} {\bibinfo  {journal} {Proceedings of the National
  Academy of Sciences}\ }\textbf {\bibinfo {volume} {102}},\ \bibinfo {pages}
  {10040} (\bibinfo {year} {2005})}\BibitemShut {NoStop}%
\bibitem [{\citenamefont {Cao}\ \emph {et~al.}(2015)\citenamefont {Cao},
  \citenamefont {Wang}, \citenamefont {Ouyang},\ and\ \citenamefont
  {Tu}}]{cao_free-energy_2015}%
  \BibitemOpen
  \bibfield  {author} {\bibinfo {author} {\bibfnamefont {Y.}~\bibnamefont
  {Cao}}, \bibinfo {author} {\bibfnamefont {H.}~\bibnamefont {Wang}}, \bibinfo
  {author} {\bibfnamefont {Q.}~\bibnamefont {Ouyang}},\ and\ \bibinfo {author}
  {\bibfnamefont {Y.}~\bibnamefont {Tu}},\ }\bibfield  {title} {\bibinfo
  {title} {The free-energy cost of accurate biochemical oscillations},\ }\href
  {https://doi.org/10.1038/nphys3412} {\bibfield  {journal} {\bibinfo
  {journal} {Nature Phys}\ }\textbf {\bibinfo {volume} {11}},\ \bibinfo {pages}
  {772} (\bibinfo {year} {2015})}\BibitemShut {NoStop}%
\bibitem [{\citenamefont {Goldbeter}\ and\ \citenamefont
  {Koshland}(1981)}]{goldbeter1981amplified}%
  \BibitemOpen
  \bibfield  {author} {\bibinfo {author} {\bibfnamefont {A.}~\bibnamefont
  {Goldbeter}}\ and\ \bibinfo {author} {\bibfnamefont {D.~E.}\ \bibnamefont
  {Koshland}},\ }\bibfield  {title} {\bibinfo {title} {An amplified sensitivity
  arising from covalent modification in biological systems.},\ }\href
  {https://doi.org/10.1073/pnas.78.11.6840} {\bibfield  {journal} {\bibinfo
  {journal} {Proceedings of the National Academy of Sciences}\ }\textbf
  {\bibinfo {volume} {78}},\ \bibinfo {pages} {6840} (\bibinfo {year}
  {1981})}\BibitemShut {NoStop}%
\bibitem [{\citenamefont {Govern}\ and\ \citenamefont {ten
  Wolde}(2014)}]{govern2014optimal}%
  \BibitemOpen
  \bibfield  {author} {\bibinfo {author} {\bibfnamefont {C.~C.}\ \bibnamefont
  {Govern}}\ and\ \bibinfo {author} {\bibfnamefont {P.~R.}\ \bibnamefont {ten
  Wolde}},\ }\bibfield  {title} {\bibinfo {title} {Optimal resource allocation
  in cellular sensing systems},\ }\href
  {https://doi.org/10.1073/pnas.1411524111} {\bibfield  {journal} {\bibinfo
  {journal} {Proceedings of the National Academy of Sciences}\ }\textbf
  {\bibinfo {volume} {111}},\ \bibinfo {pages} {17486} (\bibinfo {year}
  {2014})}\BibitemShut {NoStop}%
\bibitem [{\citenamefont {Mehta}\ \emph {et~al.}(2016)\citenamefont {Mehta},
  \citenamefont {Lang},\ and\ \citenamefont {Schwab}}]{mehta2016landauer}%
  \BibitemOpen
  \bibfield  {author} {\bibinfo {author} {\bibfnamefont {P.}~\bibnamefont
  {Mehta}}, \bibinfo {author} {\bibfnamefont {A.~H.}\ \bibnamefont {Lang}},\
  and\ \bibinfo {author} {\bibfnamefont {D.~J.}\ \bibnamefont {Schwab}},\
  }\bibfield  {title} {\bibinfo {title} {Landauer in the age of synthetic
  biology: Energy consumption and information processing in biochemical
  networks},\ }\href {https://doi.org/10.1007/s10955-015-1431-6} {\bibfield
  {journal} {\bibinfo  {journal} {Journal of Statistical Physics}\ }\textbf
  {\bibinfo {volume} {162}},\ \bibinfo {pages} {1153} (\bibinfo {year}
  {2016})}\BibitemShut {NoStop}%
\bibitem [{\citenamefont {Ouldridge}\ \emph {et~al.}(2017)\citenamefont
  {Ouldridge}, \citenamefont {Govern},\ and\ \citenamefont {ten
  Wolde}}]{ouldridge2017thermodynamics}%
  \BibitemOpen
  \bibfield  {author} {\bibinfo {author} {\bibfnamefont {T.~E.}\ \bibnamefont
  {Ouldridge}}, \bibinfo {author} {\bibfnamefont {C.~C.}\ \bibnamefont
  {Govern}},\ and\ \bibinfo {author} {\bibfnamefont {P.~R.}\ \bibnamefont {ten
  Wolde}},\ }\bibfield  {title} {\bibinfo {title} {Thermodynamics of
  computational copying in biochemical systems},\ }\href
  {https://doi.org/10.1103/PhysRevX.7.021004} {\bibfield  {journal} {\bibinfo
  {journal} {Phys. Rev. X}\ }\textbf {\bibinfo {volume} {7}},\ \bibinfo {pages}
  {021004} (\bibinfo {year} {2017})}\BibitemShut {NoStop}%
\bibitem [{\citenamefont {ten Wolde}\ \emph {et~al.}(2016)\citenamefont {ten
  Wolde}, \citenamefont {Becker}, \citenamefont {Ouldridge},\ and\
  \citenamefont {Mugler}}]{theWolde2016fundamental}%
  \BibitemOpen
  \bibfield  {author} {\bibinfo {author} {\bibfnamefont {P.~R.}\ \bibnamefont
  {ten Wolde}}, \bibinfo {author} {\bibfnamefont {N.~B.}\ \bibnamefont
  {Becker}}, \bibinfo {author} {\bibfnamefont {T.~E.}\ \bibnamefont
  {Ouldridge}},\ and\ \bibinfo {author} {\bibfnamefont {A.}~\bibnamefont
  {Mugler}},\ }\bibfield  {title} {\bibinfo {title} {Fundamental limits to
  cellular sensing},\ }\href@noop {} {\bibfield  {journal} {\bibinfo  {journal}
  {Journal of Statistical Physics}\ }\textbf {\bibinfo {volume} {162}},\
  \bibinfo {pages} {1395} (\bibinfo {year} {2016})}\BibitemShut {NoStop}%
\bibitem [{\citenamefont {Lan}\ \emph {et~al.}(2012)\citenamefont {Lan},
  \citenamefont {Sartori}, \citenamefont {Neumann}, \citenamefont {Sourjik},\
  and\ \citenamefont {Tu}}]{lan_energyspeedaccuracy_2012}%
  \BibitemOpen
  \bibfield  {author} {\bibinfo {author} {\bibfnamefont {G.}~\bibnamefont
  {Lan}}, \bibinfo {author} {\bibfnamefont {P.}~\bibnamefont {Sartori}},
  \bibinfo {author} {\bibfnamefont {S.}~\bibnamefont {Neumann}}, \bibinfo
  {author} {\bibfnamefont {V.}~\bibnamefont {Sourjik}},\ and\ \bibinfo {author}
  {\bibfnamefont {Y.}~\bibnamefont {Tu}},\ }\bibfield  {title} {\bibinfo
  {title} {The energy–speed–accuracy trade-off in sensory adaptation},\
  }\href {https://doi.org/10.1038/nphys2276} {\bibfield  {journal} {\bibinfo
  {journal} {Nature Phys}\ }\textbf {\bibinfo {volume} {8}},\ \bibinfo {pages}
  {422} (\bibinfo {year} {2012})}\BibitemShut {NoStop}%
\bibitem [{\citenamefont {Hathcock}\ \emph {et~al.}(2023)\citenamefont
  {Hathcock}, \citenamefont {Yu}, \citenamefont {Mello}, \citenamefont {Amin},
  \citenamefont {Hazelbauer},\ and\ \citenamefont
  {Tu}}]{hathcock_nonequilibrium_2023}%
  \BibitemOpen
  \bibfield  {author} {\bibinfo {author} {\bibfnamefont {D.}~\bibnamefont
  {Hathcock}}, \bibinfo {author} {\bibfnamefont {Q.}~\bibnamefont {Yu}},
  \bibinfo {author} {\bibfnamefont {B.~A.}\ \bibnamefont {Mello}}, \bibinfo
  {author} {\bibfnamefont {D.~N.}\ \bibnamefont {Amin}}, \bibinfo {author}
  {\bibfnamefont {G.~L.}\ \bibnamefont {Hazelbauer}},\ and\ \bibinfo {author}
  {\bibfnamefont {Y.}~\bibnamefont {Tu}},\ }\bibfield  {title} {\bibinfo
  {title} {A nonequilibrium allosteric model for receptor-kinase complexes:
  {The} role of energy dissipation in chemotaxis signaling},\ }\href
  {https://doi.org/10.1073/pnas.2303115120} {\bibfield  {journal} {\bibinfo
  {journal} {Proc. Natl. Acad. Sci. U.S.A.}\ }\textbf {\bibinfo {volume}
  {120}},\ \bibinfo {pages} {e2303115120} (\bibinfo {year} {2023})}\BibitemShut
  {NoStop}%
\bibitem [{\citenamefont {Tjalma}\ \emph {et~al.}(2023)\citenamefont {Tjalma},
  \citenamefont {Galstyan}, \citenamefont {Goedhart}, \citenamefont {Slim},
  \citenamefont {Becker},\ and\ \citenamefont
  {Ten~Wolde}}]{tjalma_trade-offs_2023}%
  \BibitemOpen
  \bibfield  {author} {\bibinfo {author} {\bibfnamefont {A.~J.}\ \bibnamefont
  {Tjalma}}, \bibinfo {author} {\bibfnamefont {V.}~\bibnamefont {Galstyan}},
  \bibinfo {author} {\bibfnamefont {J.}~\bibnamefont {Goedhart}}, \bibinfo
  {author} {\bibfnamefont {L.}~\bibnamefont {Slim}}, \bibinfo {author}
  {\bibfnamefont {N.~B.}\ \bibnamefont {Becker}},\ and\ \bibinfo {author}
  {\bibfnamefont {P.~R.}\ \bibnamefont {Ten~Wolde}},\ }\bibfield  {title}
  {\bibinfo {title} {Trade-offs between cost and information in cellular
  prediction},\ }\href {https://doi.org/10.1073/pnas.2303078120} {\bibfield
  {journal} {\bibinfo  {journal} {Proc. Natl. Acad. Sci. U.S.A.}\ }\textbf
  {\bibinfo {volume} {120}},\ \bibinfo {pages} {e2303078120} (\bibinfo {year}
  {2023})}\BibitemShut {NoStop}%
\bibitem [{\citenamefont {Briegel}\ \emph {et~al.}(2009)\citenamefont
  {Briegel}, \citenamefont {Ortega}, \citenamefont {Tocheva}, \citenamefont
  {Wuichet}, \citenamefont {Li}, \citenamefont {Chen}, \citenamefont
  {M{\"u}ller}, \citenamefont {Iancu}, \citenamefont {Murphy}, \citenamefont
  {Dobro} \emph {et~al.}}]{briegel2009universal}%
  \BibitemOpen
  \bibfield  {author} {\bibinfo {author} {\bibfnamefont {A.}~\bibnamefont
  {Briegel}}, \bibinfo {author} {\bibfnamefont {D.}~\bibnamefont {Ortega}},
  \bibinfo {author} {\bibfnamefont {E.}~\bibnamefont {Tocheva}}, \bibinfo
  {author} {\bibfnamefont {K.}~\bibnamefont {Wuichet}}, \bibinfo {author}
  {\bibfnamefont {Z.}~\bibnamefont {Li}}, \bibinfo {author} {\bibfnamefont
  {S.}~\bibnamefont {Chen}}, \bibinfo {author} {\bibfnamefont {A.}~\bibnamefont
  {M{\"u}ller}}, \bibinfo {author} {\bibfnamefont {C.}~\bibnamefont {Iancu}},
  \bibinfo {author} {\bibfnamefont {G.}~\bibnamefont {Murphy}}, \bibinfo
  {author} {\bibfnamefont {M.}~\bibnamefont {Dobro}}, \emph {et~al.},\
  }\bibfield  {title} {\bibinfo {title} {Universal architecture of bacterial
  chemoreceptor arrays},\ }\href@noop {} {\bibfield  {journal} {\bibinfo
  {journal} {Proc. Natl. Acad. Sci. USA}\ }\textbf {\bibinfo {volume} {106}},\
  \bibinfo {pages} {17181} (\bibinfo {year} {2009})}\BibitemShut {NoStop}%
\bibitem [{\citenamefont {Liu}\ \emph {et~al.}(2012)\citenamefont {Liu},
  \citenamefont {Hu}, \citenamefont {Morado}, \citenamefont {Jani},
  \citenamefont {Manson},\ and\ \citenamefont {Margolin}}]{liu_molecular_2012}%
  \BibitemOpen
  \bibfield  {author} {\bibinfo {author} {\bibfnamefont {J.}~\bibnamefont
  {Liu}}, \bibinfo {author} {\bibfnamefont {B.}~\bibnamefont {Hu}}, \bibinfo
  {author} {\bibfnamefont {D.~R.}\ \bibnamefont {Morado}}, \bibinfo {author}
  {\bibfnamefont {S.}~\bibnamefont {Jani}}, \bibinfo {author} {\bibfnamefont
  {M.~D.}\ \bibnamefont {Manson}},\ and\ \bibinfo {author} {\bibfnamefont
  {W.}~\bibnamefont {Margolin}},\ }\bibfield  {title} {\bibinfo {title}
  {Molecular architecture of chemoreceptor arrays revealed by cryoelectron
  tomography of \textit{{Escherichia} coli} minicells},\ }\href@noop {}
  {\bibfield  {journal} {\bibinfo  {journal} {Proc. Natl. Acad. Sci. U.S.A.}\
  }\textbf {\bibinfo {volume} {109}} (\bibinfo {year} {2012})}\BibitemShut
  {NoStop}%
\bibitem [{\citenamefont {Briegel}\ \emph {et~al.}(2012)\citenamefont
  {Briegel}, \citenamefont {Li}, \citenamefont {Bilwes}, \citenamefont
  {Hughes}, \citenamefont {Jensen},\ and\ \citenamefont
  {Crane}}]{Briegel2102Bacterial}%
  \BibitemOpen
  \bibfield  {author} {\bibinfo {author} {\bibfnamefont {A.}~\bibnamefont
  {Briegel}}, \bibinfo {author} {\bibfnamefont {X.}~\bibnamefont {Li}},
  \bibinfo {author} {\bibfnamefont {A.~M.}\ \bibnamefont {Bilwes}}, \bibinfo
  {author} {\bibfnamefont {K.~T.}\ \bibnamefont {Hughes}}, \bibinfo {author}
  {\bibfnamefont {G.~J.}\ \bibnamefont {Jensen}},\ and\ \bibinfo {author}
  {\bibfnamefont {B.~R.}\ \bibnamefont {Crane}},\ }\bibfield  {title} {\bibinfo
  {title} {Bacterial chemoreceptor arrays are hexagonally packed trimers of
  receptor dimers networked by rings of kinase and coupling proteins},\
  }\bibfield  {journal} {\bibinfo  {journal} {Proc. Natl. Acad. Sci. USA}\
  }\href {https://doi.org/10.1073/pnas.1115719109} {10.1073/pnas.1115719109}
  (\bibinfo {year} {2012})\BibitemShut {NoStop}%
\bibitem [{\citenamefont {Cassidy}\ \emph {et~al.}(2020)\citenamefont
  {Cassidy}, \citenamefont {Himes}, \citenamefont {Sun}, \citenamefont {Ma},
  \citenamefont {Zhao}, \citenamefont {Parkinson}, \citenamefont {Stansfeld},
  \citenamefont {Luthey-Schulten},\ and\ \citenamefont
  {Zhang}}]{cassidy2020structure}%
  \BibitemOpen
  \bibfield  {author} {\bibinfo {author} {\bibfnamefont {C.~K.}\ \bibnamefont
  {Cassidy}}, \bibinfo {author} {\bibfnamefont {B.~A.}\ \bibnamefont {Himes}},
  \bibinfo {author} {\bibfnamefont {D.}~\bibnamefont {Sun}}, \bibinfo {author}
  {\bibfnamefont {J.}~\bibnamefont {Ma}}, \bibinfo {author} {\bibfnamefont
  {G.}~\bibnamefont {Zhao}}, \bibinfo {author} {\bibfnamefont {J.~S.}\
  \bibnamefont {Parkinson}}, \bibinfo {author} {\bibfnamefont {P.~J.}\
  \bibnamefont {Stansfeld}}, \bibinfo {author} {\bibfnamefont {Z.}~\bibnamefont
  {Luthey-Schulten}},\ and\ \bibinfo {author} {\bibfnamefont {P.}~\bibnamefont
  {Zhang}},\ }\bibfield  {title} {\bibinfo {title} {Structure and dynamics of
  the e. coli chemotaxis core signaling complex by cryo-electron tomography and
  molecular simulations},\ }\href {https://doi.org/10.1038/s42003-019-0748-0}
  {\bibfield  {journal} {\bibinfo  {journal} {Communications Biology}\ }\textbf
  {\bibinfo {volume} {3}},\ \bibinfo {pages} {24} (\bibinfo {year}
  {2020})}\BibitemShut {NoStop}%
\bibitem [{\citenamefont {Shimizu}\ \emph {et~al.}(2010)\citenamefont
  {Shimizu}, \citenamefont {Tu},\ and\ \citenamefont
  {Berg}}]{shimizu2010modular}%
  \BibitemOpen
  \bibfield  {author} {\bibinfo {author} {\bibfnamefont {T.}~\bibnamefont
  {Shimizu}}, \bibinfo {author} {\bibfnamefont {Y.}~\bibnamefont {Tu}},\ and\
  \bibinfo {author} {\bibfnamefont {H.}~\bibnamefont {Berg}},\ }\bibfield
  {title} {\bibinfo {title} {A modular gradient-sensing network for chemotaxis
  in escherichia coli revealed by responses to time-varying stimuli},\
  }\href@noop {} {\bibfield  {journal} {\bibinfo  {journal} {Molecular systems
  biology}\ }\textbf {\bibinfo {volume} {6}} (\bibinfo {year}
  {2010})}\BibitemShut {NoStop}%
\bibitem [{\citenamefont {Amin}\ and\ \citenamefont
  {Hazelbauer}(2010)}]{Amin2010Chemoreceptors}%
  \BibitemOpen
  \bibfield  {author} {\bibinfo {author} {\bibfnamefont {D.~N.}\ \bibnamefont
  {Amin}}\ and\ \bibinfo {author} {\bibfnamefont {G.~L.}\ \bibnamefont
  {Hazelbauer}},\ }\bibfield  {title} {\bibinfo {title} {Chemoreceptors in
  signalling complexes: shifted conformation and asymmetric coupling},\ }\href
  {https://doi.org/https://doi.org/10.1111/j.1365-2958.2010.07408.x} {\bibfield
   {journal} {\bibinfo  {journal} {Molecular Microbiology}\ }\textbf {\bibinfo
  {volume} {78}},\ \bibinfo {pages} {1313} (\bibinfo {year}
  {2010})}\BibitemShut {NoStop}%
\bibitem [{\citenamefont {Tu}(2013)}]{tu2013quantitative}%
  \BibitemOpen
  \bibfield  {author} {\bibinfo {author} {\bibfnamefont {Y.}~\bibnamefont
  {Tu}},\ }\bibfield  {title} {\bibinfo {title} {Quantitative modeling of
  bacterial chemotaxis: signal amplification and accurate adaptation},\
  }\href@noop {} {\bibfield  {journal} {\bibinfo  {journal} {Annual review of
  biophysics}\ }\textbf {\bibinfo {volume} {42}},\ \bibinfo {pages} {337}
  (\bibinfo {year} {2013})}\BibitemShut {NoStop}%
\bibitem [{\citenamefont {Keegstra}\ \emph {et~al.}(2017)\citenamefont
  {Keegstra}, \citenamefont {Kamino}, \citenamefont {Anquez}, \citenamefont
  {Lazova}, \citenamefont {Emonet},\ and\ \citenamefont
  {Shimizu}}]{keegstra_phenotypic_2017}%
  \BibitemOpen
  \bibfield  {author} {\bibinfo {author} {\bibfnamefont {J.~M.}\ \bibnamefont
  {Keegstra}}, \bibinfo {author} {\bibfnamefont {K.}~\bibnamefont {Kamino}},
  \bibinfo {author} {\bibfnamefont {F.}~\bibnamefont {Anquez}}, \bibinfo
  {author} {\bibfnamefont {M.~D.}\ \bibnamefont {Lazova}}, \bibinfo {author}
  {\bibfnamefont {T.}~\bibnamefont {Emonet}},\ and\ \bibinfo {author}
  {\bibfnamefont {T.~S.}\ \bibnamefont {Shimizu}},\ }\bibfield  {title}
  {\bibinfo {title} {Phenotypic diversity and temporal variability in a
  bacterial signaling network revealed by single-cell {FRET}},\ }\href
  {https://doi.org/10.7554/eLife.27455} {\bibfield  {journal} {\bibinfo
  {journal} {eLife}\ }\textbf {\bibinfo {volume} {6}},\ \bibinfo {pages}
  {e27455} (\bibinfo {year} {2017})}\BibitemShut {NoStop}%
\bibitem [{\citenamefont {Keegstra}\ \emph {et~al.}(2022)\citenamefont
  {Keegstra}, \citenamefont {Avgidis}, \citenamefont {Mullah}, \citenamefont
  {Parkinson},\ and\ \citenamefont {Shimizu}}]{keegstra_near-critical_2022}%
  \BibitemOpen
  \bibfield  {author} {\bibinfo {author} {\bibfnamefont {J.~M.}\ \bibnamefont
  {Keegstra}}, \bibinfo {author} {\bibfnamefont {F.}~\bibnamefont {Avgidis}},
  \bibinfo {author} {\bibfnamefont {Y.}~\bibnamefont {Mullah}}, \bibinfo
  {author} {\bibfnamefont {J.~S.}\ \bibnamefont {Parkinson}},\ and\ \bibinfo
  {author} {\bibfnamefont {T.~S.}\ \bibnamefont {Shimizu}},\ }\bibfield
  {title} {\bibinfo {title} {Near-critical tuning of cooperativity revealed by
  spontaneous switching in a protein signalling array},\ }\bibfield  {journal}
  {\bibinfo  {journal} {bioRxiv}\ }\href
  {https://doi.org/10.1101/2022.12.04.518992} {10.1101/2022.12.04.518992}
  (\bibinfo {year} {2022})\BibitemShut {NoStop}%
\bibitem [{\citenamefont {Mello}\ and\ \citenamefont {Tu}(2003)}]{Mello03}%
  \BibitemOpen
  \bibfield  {author} {\bibinfo {author} {\bibfnamefont {B.~A.}\ \bibnamefont
  {Mello}}\ and\ \bibinfo {author} {\bibfnamefont {Y.}~\bibnamefont {Tu}},\
  }\bibfield  {title} {\bibinfo {title} {Perfect and near perfect adaptation in
  a model of bacterial chemotaxis},\ }\href@noop {} {\bibfield  {journal}
  {\bibinfo  {journal} {Biophys. J.}\ }\textbf {\bibinfo {volume} {84}},\
  \bibinfo {pages} {2943} (\bibinfo {year} {2003})}\BibitemShut {NoStop}%
\bibitem [{\citenamefont {Lan}\ \emph {et~al.}(2011)\citenamefont {Lan},
  \citenamefont {Schulmeister}, \citenamefont {Sourjik},\ and\ \citenamefont
  {Tu}}]{Lan2011Adapt}%
  \BibitemOpen
  \bibfield  {author} {\bibinfo {author} {\bibfnamefont {G.}~\bibnamefont
  {Lan}}, \bibinfo {author} {\bibfnamefont {S.}~\bibnamefont {Schulmeister}},
  \bibinfo {author} {\bibfnamefont {V.}~\bibnamefont {Sourjik}},\ and\ \bibinfo
  {author} {\bibfnamefont {Y.}~\bibnamefont {Tu}},\ }\bibfield  {title}
  {\bibinfo {title} {Adapt locally and act globally: strategy to maintain high
  chemoreceptor sensitivity in complex environments.},\ }\bibfield  {journal}
  {\bibinfo  {journal} {Molecular systems biology}\ }\textbf {\bibinfo {volume}
  {7}},\ \href {https://doi.org/10.1038/msb.2011.8} {10.1038/msb.2011.8}
  (\bibinfo {year} {2011})\BibitemShut {NoStop}%
\bibitem [{\citenamefont {Sartori}\ and\ \citenamefont
  {Tu}(2015)}]{sartori_free_2015}%
  \BibitemOpen
  \bibfield  {author} {\bibinfo {author} {\bibfnamefont {P.}~\bibnamefont
  {Sartori}}\ and\ \bibinfo {author} {\bibfnamefont {Y.}~\bibnamefont {Tu}},\
  }\bibfield  {title} {\bibinfo {title} {Free {Energy} {Cost} of {Reducing}
  {Noise} while {Maintaining} a {High} {Sensitivity}},\ }\href
  {https://doi.org/10.1103/PhysRevLett.115.118102} {\bibfield  {journal}
  {\bibinfo  {journal} {Phys. Rev. Lett.}\ }\textbf {\bibinfo {volume} {115}},\
  \bibinfo {pages} {118102} (\bibinfo {year} {2015})}\BibitemShut {NoStop}%
\bibitem [{\citenamefont {Fei}\ \emph {et~al.}(2018)\citenamefont {Fei},
  \citenamefont {Cao}, \citenamefont {Ouyang},\ and\ \citenamefont
  {Tu}}]{fei_design_2018}%
  \BibitemOpen
  \bibfield  {author} {\bibinfo {author} {\bibfnamefont {C.}~\bibnamefont
  {Fei}}, \bibinfo {author} {\bibfnamefont {Y.}~\bibnamefont {Cao}}, \bibinfo
  {author} {\bibfnamefont {Q.}~\bibnamefont {Ouyang}},\ and\ \bibinfo {author}
  {\bibfnamefont {Y.}~\bibnamefont {Tu}},\ }\bibfield  {title} {\bibinfo
  {title} {Design principles for enhancing phase sensitivity and suppressing
  phase fluctuations simultaneously in biochemical oscillatory systems},\
  }\href {https://doi.org/10.1038/s41467-018-03826-4} {\bibfield  {journal}
  {\bibinfo  {journal} {Nat Commun}\ }\textbf {\bibinfo {volume} {9}},\
  \bibinfo {pages} {1434} (\bibinfo {year} {2018})}\BibitemShut {NoStop}%
\bibitem [{\citenamefont {Parkinson}\ \emph {et~al.}(2015)\citenamefont
  {Parkinson}, \citenamefont {Hazelbauer},\ and\ \citenamefont
  {Falke}}]{parkinson_signaling_2015}%
  \BibitemOpen
  \bibfield  {author} {\bibinfo {author} {\bibfnamefont {J.~S.}\ \bibnamefont
  {Parkinson}}, \bibinfo {author} {\bibfnamefont {G.~L.}\ \bibnamefont
  {Hazelbauer}},\ and\ \bibinfo {author} {\bibfnamefont {J.~J.}\ \bibnamefont
  {Falke}},\ }\bibfield  {title} {\bibinfo {title} {Signaling and sensory
  adaptation in {Escherichia} coli chemoreceptors: 2015 update},\ }\href
  {https://doi.org/10.1016/j.tim.2015.03.003} {\bibfield  {journal} {\bibinfo
  {journal} {Trends in Microbiology}\ }\textbf {\bibinfo {volume} {23}},\
  \bibinfo {pages} {257} (\bibinfo {year} {2015})}\BibitemShut {NoStop}%
\bibitem [{\citenamefont {Levit}\ and\ \citenamefont {Stock}(2002)}]{Levit02}%
  \BibitemOpen
  \bibfield  {author} {\bibinfo {author} {\bibfnamefont {M.~N.}\ \bibnamefont
  {Levit}}\ and\ \bibinfo {author} {\bibfnamefont {J.~B.}\ \bibnamefont
  {Stock}},\ }\bibfield  {title} {\bibinfo {title} {Receptor methylation
  controls the magnitude of stimulus-response coupling in bacterial
  chemotaxis},\ }\href@noop {} {\bibfield  {journal} {\bibinfo  {journal} {J.
  Biol. Chem.}\ }\textbf {\bibinfo {volume} {277}},\ \bibinfo {pages} {36760}
  (\bibinfo {year} {2002})}\BibitemShut {NoStop}%
\bibitem [{\citenamefont {Vaknin}\ and\ \citenamefont {Berg}(2007)}]{Vaknin07}%
  \BibitemOpen
  \bibfield  {author} {\bibinfo {author} {\bibfnamefont {A.}~\bibnamefont
  {Vaknin}}\ and\ \bibinfo {author} {\bibfnamefont {H.~C.}\ \bibnamefont
  {Berg}},\ }\bibfield  {title} {\bibinfo {title} {Physical responses of
  bacterial chemoreceptors.},\ }\href@noop {} {\bibfield  {journal} {\bibinfo
  {journal} {J. Mol. Biol.}\ }\textbf {\bibinfo {volume} {366}},\ \bibinfo
  {pages} {1416} (\bibinfo {year} {2007})}\BibitemShut {NoStop}%
\bibitem [{\citenamefont {Monod}\ \emph {et~al.}(1965)\citenamefont {Monod},
  \citenamefont {Wyman},\ and\ \citenamefont {Changeux}}]{MWC1965}%
  \BibitemOpen
  \bibfield  {author} {\bibinfo {author} {\bibfnamefont {J.}~\bibnamefont
  {Monod}}, \bibinfo {author} {\bibfnamefont {J.}~\bibnamefont {Wyman}},\ and\
  \bibinfo {author} {\bibfnamefont {J.}~\bibnamefont {Changeux}},\ }\bibfield
  {title} {\bibinfo {title} {{On the nature of allosteric transitions: a
  plausible model.}},\ }\href@noop {} {\bibfield  {journal} {\bibinfo
  {journal} {Journal of molecular biology}\ }\textbf {\bibinfo {volume} {12}},\
  \bibinfo {pages} {88} (\bibinfo {year} {1965})}\BibitemShut {NoStop}%
\bibitem [{\citenamefont {Li}\ and\ \citenamefont
  {Hazelbauer}(2014)}]{li2014selective}%
  \BibitemOpen
  \bibfield  {author} {\bibinfo {author} {\bibfnamefont {M.}~\bibnamefont
  {Li}}\ and\ \bibinfo {author} {\bibfnamefont {G.~L.}\ \bibnamefont
  {Hazelbauer}},\ }\bibfield  {title} {\bibinfo {title} {Selective allosteric
  coupling in core chemotaxis signaling complexes},\ }\href
  {https://doi.org/10.1073/pnas.1415184111} {\bibfield  {journal} {\bibinfo
  {journal} {Proceedings of the National Academy of Sciences}\ }\textbf
  {\bibinfo {volume} {111}},\ \bibinfo {pages} {15940} (\bibinfo {year}
  {2014})}\BibitemShut {NoStop}%
\bibitem [{\citenamefont {Piñas}\ \emph {et~al.}(2016)\citenamefont {Piñas},
  \citenamefont {Frank}, \citenamefont {Vaknin},\ and\ \citenamefont
  {Parkinson}}]{pinas2016source}%
  \BibitemOpen
  \bibfield  {author} {\bibinfo {author} {\bibfnamefont {G.~E.}\ \bibnamefont
  {Piñas}}, \bibinfo {author} {\bibfnamefont {V.}~\bibnamefont {Frank}},
  \bibinfo {author} {\bibfnamefont {A.}~\bibnamefont {Vaknin}},\ and\ \bibinfo
  {author} {\bibfnamefont {J.~S.}\ \bibnamefont {Parkinson}},\ }\bibfield
  {title} {\bibinfo {title} {The source of high signal cooperativity in
  bacterial chemosensory arrays},\ }\href
  {https://doi.org/10.1073/pnas.1600216113} {\bibfield  {journal} {\bibinfo
  {journal} {Proceedings of the National Academy of Sciences}\ }\textbf
  {\bibinfo {volume} {113}},\ \bibinfo {pages} {3335} (\bibinfo {year}
  {2016})}\BibitemShut {NoStop}%
\bibitem [{\citenamefont {Gillespie}(1977)}]{gillespie_exact_1977}%
  \BibitemOpen
  \bibfield  {author} {\bibinfo {author} {\bibfnamefont {D.~T.}\ \bibnamefont
  {Gillespie}},\ }\bibfield  {title} {\bibinfo {title} {Exact stochastic
  simulation of coupled chemical reactions},\ }\href
  {https://doi.org/10.1021/j100540a008} {\bibfield  {journal} {\bibinfo
  {journal} {J. Phys. Chem.}\ }\textbf {\bibinfo {volume} {81}},\ \bibinfo
  {pages} {2340} (\bibinfo {year} {1977})}\BibitemShut {NoStop}%
\bibitem [{\citenamefont {E}\ and\ \citenamefont
  {Vanden-Eijnden}(2010)}]{e_transition-path_2010}%
  \BibitemOpen
  \bibfield  {author} {\bibinfo {author} {\bibfnamefont {W.}~\bibnamefont {E}}\
  and\ \bibinfo {author} {\bibfnamefont {E.}~\bibnamefont {Vanden-Eijnden}},\
  }\bibfield  {title} {\bibinfo {title} {Transition-{Path} {Theory} and
  {Path}-{Finding} {Algorithms} for the {Study} of {Rare} {Events}},\ }\href
  {https://doi.org/10.1146/annurev.physchem.040808.090412} {\bibfield
  {journal} {\bibinfo  {journal} {Annual Review of Physical Chemistry}\
  }\textbf {\bibinfo {volume} {61}},\ \bibinfo {pages} {391} (\bibinfo {year}
  {2010})}\BibitemShut {NoStop}%
\bibitem [{\citenamefont {Berezhkovskii}\ and\ \citenamefont
  {Makarov}(2019)}]{berezhkovskii_forwardbackward_2019}%
  \BibitemOpen
  \bibfield  {author} {\bibinfo {author} {\bibfnamefont {A.~M.}\ \bibnamefont
  {Berezhkovskii}}\ and\ \bibinfo {author} {\bibfnamefont {D.~E.}\ \bibnamefont
  {Makarov}},\ }\bibfield  {title} {\bibinfo {title} {On the forward/backward
  symmetry of transition path time distributions in nonequilibrium systems},\
  }\href {https://doi.org/10.1063/1.5109293} {\bibfield  {journal} {\bibinfo
  {journal} {J. Chem. Phys.}\ }\textbf {\bibinfo {volume} {151}},\ \bibinfo
  {pages} {065102} (\bibinfo {year} {2019})}\BibitemShut {NoStop}%
\bibitem [{\citenamefont {Sourjik}\ and\ \citenamefont
  {Berg}(2002)}]{Sourjik02a}%
  \BibitemOpen
  \bibfield  {author} {\bibinfo {author} {\bibfnamefont {V.}~\bibnamefont
  {Sourjik}}\ and\ \bibinfo {author} {\bibfnamefont {H.~C.}\ \bibnamefont
  {Berg}},\ }\bibfield  {title} {\bibinfo {title} {Binding of the {{\it E.
  coli}} response regulator {CheY} to its target is measured in vivo by
  fluorescence resonance energy transfer},\ }\href@noop {} {\bibfield
  {journal} {\bibinfo  {journal} {Proc. Natl. Acad. Sci. USA}\ }\textbf
  {\bibinfo {volume} {99}},\ \bibinfo {pages} {12669} (\bibinfo {year}
  {2002})}\BibitemShut {NoStop}%
\bibitem [{\citenamefont {Yu}\ \emph {et~al.}(2021)\citenamefont {Yu},
  \citenamefont {Zhang},\ and\ \citenamefont {Tu}}]{yu2021inverse}%
  \BibitemOpen
  \bibfield  {author} {\bibinfo {author} {\bibfnamefont {Q.}~\bibnamefont
  {Yu}}, \bibinfo {author} {\bibfnamefont {D.}~\bibnamefont {Zhang}},\ and\
  \bibinfo {author} {\bibfnamefont {Y.}~\bibnamefont {Tu}},\ }\bibfield
  {title} {\bibinfo {title} {Inverse power law scaling of energy dissipation
  rate in nonequilibrium reaction networks},\ }\href
  {https://doi.org/10.1103/PhysRevLett.126.080601} {\bibfield  {journal}
  {\bibinfo  {journal} {Phys. Rev. Lett.}\ }\textbf {\bibinfo {volume} {126}},\
  \bibinfo {pages} {080601} (\bibinfo {year} {2021})}\BibitemShut {NoStop}%
\bibitem [{\citenamefont {Yu}\ and\ \citenamefont
  {Tu}(2022{\natexlab{a}})}]{yu_state-space_2022}%
  \BibitemOpen
  \bibfield  {author} {\bibinfo {author} {\bibfnamefont {Q.}~\bibnamefont
  {Yu}}\ and\ \bibinfo {author} {\bibfnamefont {Y.}~\bibnamefont {Tu}},\
  }\bibfield  {title} {\bibinfo {title} {State-space renormalization group
  theory of nonequilibrium reaction networks: {Exact} solutions for hypercubic
  lattices in arbitrary dimensions},\ }\href
  {https://doi.org/10.1103/PhysRevE.105.044140} {\bibfield  {journal} {\bibinfo
   {journal} {Phys. Rev. E}\ }\textbf {\bibinfo {volume} {105}},\ \bibinfo
  {pages} {044140} (\bibinfo {year} {2022}{\natexlab{a}})}\BibitemShut
  {NoStop}%
\bibitem [{\citenamefont {Hänggi}\ \emph {et~al.}(1990)\citenamefont
  {Hänggi}, \citenamefont {Talkner},\ and\ \citenamefont
  {Borkovec}}]{hanggi_reaction-rate_1990}%
  \BibitemOpen
  \bibfield  {author} {\bibinfo {author} {\bibfnamefont {P.}~\bibnamefont
  {Hänggi}}, \bibinfo {author} {\bibfnamefont {P.}~\bibnamefont {Talkner}},\
  and\ \bibinfo {author} {\bibfnamefont {M.}~\bibnamefont {Borkovec}},\
  }\bibfield  {title} {\bibinfo {title} {Reaction-rate theory: fifty years
  after {Kramers}},\ }\href {https://doi.org/10.1103/RevModPhys.62.251}
  {\bibfield  {journal} {\bibinfo  {journal} {Rev. Mod. Phys.}\ }\textbf
  {\bibinfo {volume} {62}},\ \bibinfo {pages} {251} (\bibinfo {year}
  {1990})}\BibitemShut {NoStop}%
\bibitem [{\citenamefont {Hummer}(2004)}]{hummer_transition_2004}%
  \BibitemOpen
  \bibfield  {author} {\bibinfo {author} {\bibfnamefont {G.}~\bibnamefont
  {Hummer}},\ }\bibfield  {title} {\bibinfo {title} {From transition paths to
  transition states and rate coefficients},\ }\href
  {https://doi.org/10.1063/1.1630572} {\bibfield  {journal} {\bibinfo
  {journal} {The Journal of Chemical Physics}\ }\textbf {\bibinfo {volume}
  {120}},\ \bibinfo {pages} {516} (\bibinfo {year} {2004})}\BibitemShut
  {NoStop}%
\bibitem [{\citenamefont {Chung}\ \emph {et~al.}(2009)\citenamefont {Chung},
  \citenamefont {Louis},\ and\ \citenamefont
  {Eaton}}]{chung_experimental_2009}%
  \BibitemOpen
  \bibfield  {author} {\bibinfo {author} {\bibfnamefont {H.~S.}\ \bibnamefont
  {Chung}}, \bibinfo {author} {\bibfnamefont {J.~M.}\ \bibnamefont {Louis}},\
  and\ \bibinfo {author} {\bibfnamefont {W.~A.}\ \bibnamefont {Eaton}},\
  }\bibfield  {title} {\bibinfo {title} {Experimental determination of upper
  bound for transition path times in protein folding from single-molecule
  photon-by-photon trajectories},\ }\href
  {https://doi.org/10.1073/pnas.0901178106} {\bibfield  {journal} {\bibinfo
  {journal} {Proc. Natl. Acad. Sci. U.S.A.}\ }\textbf {\bibinfo {volume}
  {106}},\ \bibinfo {pages} {11837} (\bibinfo {year} {2009})}\BibitemShut
  {NoStop}%
\bibitem [{\citenamefont {Tu}(2008)}]{tu2008nonequilibrium}%
  \BibitemOpen
  \bibfield  {author} {\bibinfo {author} {\bibfnamefont {Y.}~\bibnamefont
  {Tu}},\ }\bibfield  {title} {\bibinfo {title} {The nonequilibrium mechanism
  for ultrasensitivity in a biological switch: Sensing by maxwell's demons},\
  }\href {https://doi.org/10.1073/pnas.0804641105} {\bibfield  {journal}
  {\bibinfo  {journal} {Proceedings of the National Academy of Sciences}\
  }\textbf {\bibinfo {volume} {105}},\ \bibinfo {pages} {11737} (\bibinfo
  {year} {2008})}\BibitemShut {NoStop}%
\bibitem [{\citenamefont {Yu}\ and\ \citenamefont
  {Tu}(2022{\natexlab{b}})}]{yu_flocking_2022}%
  \BibitemOpen
  \bibfield  {author} {\bibinfo {author} {\bibfnamefont {Q.}~\bibnamefont
  {Yu}}\ and\ \bibinfo {author} {\bibfnamefont {Y.}~\bibnamefont {Tu}},\
  }\bibfield  {title} {\bibinfo {title} {Energy {Cost} for {Flocking} of
  {Active} {Spins}: {The} {Cusped} {Dissipation} {Maximum} at the {Flocking}
  {Transition}},\ }\href {https://doi.org/10.1103/PhysRevLett.129.278001}
  {\bibfield  {journal} {\bibinfo  {journal} {Phys. Rev. Lett.}\ }\textbf
  {\bibinfo {volume} {129}},\ \bibinfo {pages} {278001} (\bibinfo {year}
  {2022}{\natexlab{b}})}\BibitemShut {NoStop}%
\bibitem [{\citenamefont {Martin}\ \emph {et~al.}(2001)\citenamefont {Martin},
  \citenamefont {Hudspeth},\ and\ \citenamefont
  {Jülicher}}]{martin_comparison_2001}%
  \BibitemOpen
  \bibfield  {author} {\bibinfo {author} {\bibfnamefont {P.}~\bibnamefont
  {Martin}}, \bibinfo {author} {\bibfnamefont {A.~J.}\ \bibnamefont
  {Hudspeth}},\ and\ \bibinfo {author} {\bibfnamefont {F.}~\bibnamefont
  {Jülicher}},\ }\bibfield  {title} {\bibinfo {title} {Comparison of a hair
  bundle's spontaneous oscillations with its response to mechanical stimulation
  reveals the underlying active process},\ }\href
  {https://doi.org/10.1073/pnas.251530598} {\bibfield  {journal} {\bibinfo
  {journal} {Proc. Natl. Acad. Sci. U.S.A.}\ }\textbf {\bibinfo {volume}
  {98}},\ \bibinfo {pages} {14380} (\bibinfo {year} {2001})}\BibitemShut
  {NoStop}%
\bibitem [{\citenamefont {Lasanta}\ \emph {et~al.}(2017)\citenamefont
  {Lasanta}, \citenamefont {Vega~Reyes}, \citenamefont {Prados},\ and\
  \citenamefont {Santos}}]{lasanta_when_2017}%
  \BibitemOpen
  \bibfield  {author} {\bibinfo {author} {\bibfnamefont {A.}~\bibnamefont
  {Lasanta}}, \bibinfo {author} {\bibfnamefont {F.}~\bibnamefont {Vega~Reyes}},
  \bibinfo {author} {\bibfnamefont {A.}~\bibnamefont {Prados}},\ and\ \bibinfo
  {author} {\bibfnamefont {A.}~\bibnamefont {Santos}},\ }\bibfield  {title}
  {\bibinfo {title} {When the {Hotter} {Cools} {More} {Quickly}: {Mpemba}
  {Effect} in {Granular} {Fluids}},\ }\href
  {https://doi.org/10.1103/PhysRevLett.119.148001} {\bibfield  {journal}
  {\bibinfo  {journal} {Phys. Rev. Lett.}\ }\textbf {\bibinfo {volume} {119}},\
  \bibinfo {pages} {148001} (\bibinfo {year} {2017})}\BibitemShut {NoStop}%
\bibitem [{\citenamefont {Koler}\ \emph {et~al.}(2024)\citenamefont {Koler},
  \citenamefont {Parkinson},\ and\ \citenamefont {Vaknin}}]{koler2024signal}%
  \BibitemOpen
  \bibfield  {author} {\bibinfo {author} {\bibfnamefont {M.}~\bibnamefont
  {Koler}}, \bibinfo {author} {\bibfnamefont {J.~S.}\ \bibnamefont
  {Parkinson}},\ and\ \bibinfo {author} {\bibfnamefont {A.}~\bibnamefont
  {Vaknin}},\ }\bibfield  {title} {\bibinfo {title} {Signal integration in
  chemoreceptor complexes},\ }\href {https://doi.org/10.1073/pnas.2312064121}
  {\bibfield  {journal} {\bibinfo  {journal} {Proceedings of the National
  Academy of Sciences}\ }\textbf {\bibinfo {volume} {121}},\ \bibinfo {pages}
  {e2312064121} (\bibinfo {year} {2024})}\BibitemShut {NoStop}%
\bibitem [{\citenamefont {Muok}\ \emph
  {et~al.}(2020{\natexlab{a}})\citenamefont {Muok}, \citenamefont {Ortega},
  \citenamefont {Kurniyati}, \citenamefont {Yang}, \citenamefont {Maschmann},
  \citenamefont {Sidi~Mabrouk}, \citenamefont {Li}, \citenamefont {Crane},\
  and\ \citenamefont {Briegel}}]{muok2020atypical}%
  \BibitemOpen
  \bibfield  {author} {\bibinfo {author} {\bibfnamefont {A.~R.}\ \bibnamefont
  {Muok}}, \bibinfo {author} {\bibfnamefont {D.~R.}\ \bibnamefont {Ortega}},
  \bibinfo {author} {\bibfnamefont {K.}~\bibnamefont {Kurniyati}}, \bibinfo
  {author} {\bibfnamefont {W.}~\bibnamefont {Yang}}, \bibinfo {author}
  {\bibfnamefont {Z.~A.}\ \bibnamefont {Maschmann}}, \bibinfo {author}
  {\bibfnamefont {A.}~\bibnamefont {Sidi~Mabrouk}}, \bibinfo {author}
  {\bibfnamefont {C.}~\bibnamefont {Li}}, \bibinfo {author} {\bibfnamefont
  {B.~R.}\ \bibnamefont {Crane}},\ and\ \bibinfo {author} {\bibfnamefont
  {A.}~\bibnamefont {Briegel}},\ }\bibfield  {title} {\bibinfo {title}
  {Atypical chemoreceptor arrays accommodate high membrane curvature},\
  }\href@noop {} {\bibfield  {journal} {\bibinfo  {journal} {Nature
  Communications}\ }\textbf {\bibinfo {volume} {11}},\ \bibinfo {pages} {5763}
  (\bibinfo {year} {2020}{\natexlab{a}})}\BibitemShut {NoStop}%
\bibitem [{\citenamefont {Burt}\ \emph {et~al.}(2021)\citenamefont {Burt},
  \citenamefont {Cassidy}, \citenamefont {Stansfeld},\ and\ \citenamefont
  {Gutsche}}]{burt2021alternative}%
  \BibitemOpen
  \bibfield  {author} {\bibinfo {author} {\bibfnamefont {A.}~\bibnamefont
  {Burt}}, \bibinfo {author} {\bibfnamefont {C.~K.}\ \bibnamefont {Cassidy}},
  \bibinfo {author} {\bibfnamefont {P.~J.}\ \bibnamefont {Stansfeld}},\ and\
  \bibinfo {author} {\bibfnamefont {I.}~\bibnamefont {Gutsche}},\ }\bibfield
  {title} {\bibinfo {title} {Alternative architecture of the e. coli
  chemosensory array},\ }\bibfield  {journal} {\bibinfo  {journal}
  {Biomolecules}\ }\textbf {\bibinfo {volume} {11}},\ \href
  {https://doi.org/10.3390/biom11040495} {10.3390/biom11040495} (\bibinfo
  {year} {2021})\BibitemShut {NoStop}%
\bibitem [{\citenamefont {Piñas}\ \emph {et~al.}(2022)\citenamefont {Piñas},
  \citenamefont {DeSantis}, \citenamefont {Cassidy},\ and\ \citenamefont
  {Parkinson}}]{pinas2022hexameric}%
  \BibitemOpen
  \bibfield  {author} {\bibinfo {author} {\bibfnamefont {G.~E.}\ \bibnamefont
  {Piñas}}, \bibinfo {author} {\bibfnamefont {M.~D.}\ \bibnamefont
  {DeSantis}}, \bibinfo {author} {\bibfnamefont {C.~K.}\ \bibnamefont
  {Cassidy}},\ and\ \bibinfo {author} {\bibfnamefont {J.~S.}\ \bibnamefont
  {Parkinson}},\ }\bibfield  {title} {\bibinfo {title} {Hexameric rings of the
  scaffolding protein chew enhance response sensitivity and cooperativity in
  \emph{Escherichia coli} chemoreceptor arrays},\ }\href
  {https://doi.org/10.1126/scisignal.abj1737} {\bibfield  {journal} {\bibinfo
  {journal} {Science Signaling}\ }\textbf {\bibinfo {volume} {15}},\ \bibinfo
  {pages} {eabj1737} (\bibinfo {year} {2022})}\BibitemShut {NoStop}%
\bibitem [{\citenamefont {Khursigara}\ \emph {et~al.}(2011)\citenamefont
  {Khursigara}, \citenamefont {Lan}, \citenamefont {Neumann}, \citenamefont
  {Wu}, \citenamefont {Ravindran}, \citenamefont {Borgnia}, \citenamefont
  {Sourjik}, \citenamefont {Milne}, \citenamefont {Tu},\ and\ \citenamefont
  {Subramaniam}}]{khursigara2011lateral}%
  \BibitemOpen
  \bibfield  {author} {\bibinfo {author} {\bibfnamefont {C.}~\bibnamefont
  {Khursigara}}, \bibinfo {author} {\bibfnamefont {G.}~\bibnamefont {Lan}},
  \bibinfo {author} {\bibfnamefont {S.}~\bibnamefont {Neumann}}, \bibinfo
  {author} {\bibfnamefont {X.}~\bibnamefont {Wu}}, \bibinfo {author}
  {\bibfnamefont {S.}~\bibnamefont {Ravindran}}, \bibinfo {author}
  {\bibfnamefont {M.}~\bibnamefont {Borgnia}}, \bibinfo {author} {\bibfnamefont
  {V.}~\bibnamefont {Sourjik}}, \bibinfo {author} {\bibfnamefont
  {J.}~\bibnamefont {Milne}}, \bibinfo {author} {\bibfnamefont
  {Y.}~\bibnamefont {Tu}},\ and\ \bibinfo {author} {\bibfnamefont
  {S.}~\bibnamefont {Subramaniam}},\ }\bibfield  {title} {\bibinfo {title}
  {Lateral density of receptor arrays in the membrane plane influences
  sensitivity of the e. coli chemotaxis response},\ }\href@noop {} {\bibfield
  {journal} {\bibinfo  {journal} {The EMBO Journal}\ }\textbf {\bibinfo
  {volume} {30}},\ \bibinfo {pages} {1719} (\bibinfo {year}
  {2011})}\BibitemShut {NoStop}%
\bibitem [{\citenamefont {Frank}\ and\ \citenamefont
  {Vaknin}(2013)}]{frank_prolonged_2013}%
  \BibitemOpen
  \bibfield  {author} {\bibinfo {author} {\bibfnamefont {V.}~\bibnamefont
  {Frank}}\ and\ \bibinfo {author} {\bibfnamefont {A.}~\bibnamefont {Vaknin}},\
  }\bibfield  {title} {\bibinfo {title} {Prolonged stimuli alter the bacterial
  chemosensory clusters: {Stimuli} alter chemosensory clusters},\ }\href
  {https://doi.org/10.1111/mmi.12215} {\bibfield  {journal} {\bibinfo
  {journal} {Molecular Microbiology}\ }\textbf {\bibinfo {volume} {88}},\
  \bibinfo {pages} {634} (\bibinfo {year} {2013})}\BibitemShut {NoStop}%
\bibitem [{\citenamefont {Mello}\ \emph {et~al.}(2018)\citenamefont {Mello},
  \citenamefont {Pan}, \citenamefont {Hazelbauer},\ and\ \citenamefont
  {Tu}}]{mello_dual_2018}%
  \BibitemOpen
  \bibfield  {author} {\bibinfo {author} {\bibfnamefont {B.~A.}\ \bibnamefont
  {Mello}}, \bibinfo {author} {\bibfnamefont {W.}~\bibnamefont {Pan}}, \bibinfo
  {author} {\bibfnamefont {G.~L.}\ \bibnamefont {Hazelbauer}},\ and\ \bibinfo
  {author} {\bibfnamefont {Y.}~\bibnamefont {Tu}},\ }\bibfield  {title}
  {\bibinfo {title} {A dual regulation mechanism of histidine kinase {CheA}
  identified by combining network-dynamics modeling and system-level
  input-output data},\ }\href {https://doi.org/10.1371/journal.pcbi.1006305}
  {\bibfield  {journal} {\bibinfo  {journal} {PLoS Comput Biol}\ }\textbf
  {\bibinfo {volume} {14}},\ \bibinfo {pages} {e1006305} (\bibinfo {year}
  {2018})}\BibitemShut {NoStop}%
\bibitem [{\citenamefont {Muok}\ \emph
  {et~al.}(2020{\natexlab{b}})\citenamefont {Muok}, \citenamefont {Briegel},\
  and\ \citenamefont {Crane}}]{muok_regulation_2020}%
  \BibitemOpen
  \bibfield  {author} {\bibinfo {author} {\bibfnamefont {A.~R.}\ \bibnamefont
  {Muok}}, \bibinfo {author} {\bibfnamefont {A.}~\bibnamefont {Briegel}},\ and\
  \bibinfo {author} {\bibfnamefont {B.~R.}\ \bibnamefont {Crane}},\ }\bibfield
  {title} {\bibinfo {title} {Regulation of the chemotaxis histidine kinase
  {CheA}: {A} structural perspective},\ }\href
  {https://doi.org/10.1016/j.bbamem.2019.183030} {\bibfield  {journal}
  {\bibinfo  {journal} {Biochimica et Biophysica Acta (BBA) - Biomembranes}\
  }\textbf {\bibinfo {volume} {1862}},\ \bibinfo {pages} {183030} (\bibinfo
  {year} {2020}{\natexlab{b}})}\BibitemShut {NoStop}%
\bibitem [{\citenamefont {Schreiber}\ \emph {et~al.}(1983)\citenamefont
  {Schreiber}, \citenamefont {Libermann}, \citenamefont {Lax}, \citenamefont
  {Yarden},\ and\ \citenamefont {Schlessinger}}]{schreiber_biological_1983}%
  \BibitemOpen
  \bibfield  {author} {\bibinfo {author} {\bibfnamefont {A.~B.}\ \bibnamefont
  {Schreiber}}, \bibinfo {author} {\bibfnamefont {T.~A.}\ \bibnamefont
  {Libermann}}, \bibinfo {author} {\bibfnamefont {I.}~\bibnamefont {Lax}},
  \bibinfo {author} {\bibfnamefont {Y.}~\bibnamefont {Yarden}},\ and\ \bibinfo
  {author} {\bibfnamefont {J.}~\bibnamefont {Schlessinger}},\ }\bibfield
  {title} {\bibinfo {title} {Biological role of epidermal growth
  factor-receptor clustering. {Investigation} with monoclonal anti-receptor
  antibodies.},\ }\href {https://doi.org/10.1016/S0021-9258(18)33127-2}
  {\bibfield  {journal} {\bibinfo  {journal} {Journal of Biological Chemistry}\
  }\textbf {\bibinfo {volume} {258}},\ \bibinfo {pages} {846} (\bibinfo {year}
  {1983})}\BibitemShut {NoStop}%
\bibitem [{\citenamefont {S{\'a}nchez}\ and\ \citenamefont
  {Tamp{\'e}}(2023)}]{sanchez2023ligand}%
  \BibitemOpen
  \bibfield  {author} {\bibinfo {author} {\bibfnamefont {M.~F.}\ \bibnamefont
  {S{\'a}nchez}}\ and\ \bibinfo {author} {\bibfnamefont {R.}~\bibnamefont
  {Tamp{\'e}}},\ }\bibfield  {title} {\bibinfo {title} {Ligand-independent
  receptor clustering modulates transmembrane signaling: a new paradigm},\
  }\href {https://doi.org/10.1016/j.tibs.2022.08.002} {\bibfield  {journal}
  {\bibinfo  {journal} {Trends in Biochemical Sciences}\ }\textbf {\bibinfo
  {volume} {48}},\ \bibinfo {pages} {156} (\bibinfo {year} {2023})}\BibitemShut
  {NoStop}%
\bibitem [{\citenamefont {Honerkamp-Smith}\ \emph {et~al.}(2012)\citenamefont
  {Honerkamp-Smith}, \citenamefont {Machta},\ and\ \citenamefont
  {Keller}}]{honerkamp-smith_experimental_2012}%
  \BibitemOpen
  \bibfield  {author} {\bibinfo {author} {\bibfnamefont {A.~R.}\ \bibnamefont
  {Honerkamp-Smith}}, \bibinfo {author} {\bibfnamefont {B.~B.}\ \bibnamefont
  {Machta}},\ and\ \bibinfo {author} {\bibfnamefont {S.~L.}\ \bibnamefont
  {Keller}},\ }\bibfield  {title} {\bibinfo {title} {Experimental
  {Observations} of {Dynamic} {Critical} {Phenomena} in a {Lipid} {Membrane}},\
  }\href {https://doi.org/10.1103/PhysRevLett.108.265702} {\bibfield  {journal}
  {\bibinfo  {journal} {Phys. Rev. Lett.}\ }\textbf {\bibinfo {volume} {108}},\
  \bibinfo {pages} {265702} (\bibinfo {year} {2012})}\BibitemShut {NoStop}%
\bibitem [{\citenamefont {Tan}\ \emph {et~al.}(2021)\citenamefont {Tan},
  \citenamefont {Watson}, \citenamefont {Chao}, \citenamefont {Li},
  \citenamefont {Gingrich}, \citenamefont {Horowitz},\ and\ \citenamefont
  {Fakhri}}]{tan_scale-dependent_2021}%
  \BibitemOpen
  \bibfield  {author} {\bibinfo {author} {\bibfnamefont {T.~H.}\ \bibnamefont
  {Tan}}, \bibinfo {author} {\bibfnamefont {G.~A.}\ \bibnamefont {Watson}},
  \bibinfo {author} {\bibfnamefont {Y.-C.}\ \bibnamefont {Chao}}, \bibinfo
  {author} {\bibfnamefont {J.}~\bibnamefont {Li}}, \bibinfo {author}
  {\bibfnamefont {T.~R.}\ \bibnamefont {Gingrich}}, \bibinfo {author}
  {\bibfnamefont {J.~M.}\ \bibnamefont {Horowitz}},\ and\ \bibinfo {author}
  {\bibfnamefont {N.}~\bibnamefont {Fakhri}},\ }\href
  {http://arxiv.org/abs/2107.05701} {\bibinfo {title} {Scale-dependent
  irreversibility in living matter}} (\bibinfo {year} {2021}),\ \bibinfo {note}
  {arXiv:2107.05701 [cond-mat, physics:physics]}\BibitemShut {NoStop}%
\bibitem [{\citenamefont {Berg}(2003)}]{HBerg2003}%
  \BibitemOpen
  \bibfield  {author} {\bibinfo {author} {\bibfnamefont {H.~C.}\ \bibnamefont
  {Berg}},\ }\bibfield  {title} {\bibinfo {title} {The rotatory motor of
  bacterial flagella},\ }\href@noop {} {\bibfield  {journal} {\bibinfo
  {journal} {Annu. Rev. Biochem.}\ }\textbf {\bibinfo {volume} {72}},\ \bibinfo
  {pages} {19} (\bibinfo {year} {2003})}\BibitemShut {NoStop}%
\bibitem [{\citenamefont {Han}\ \emph {et~al.}(2023)\citenamefont {Han},
  \citenamefont {Zhang}, \citenamefont {Hong}, \citenamefont {Yu},
  \citenamefont {Wu}, \citenamefont {Yang}, \citenamefont {Rust}, \citenamefont
  {Tu},\ and\ \citenamefont {Ouyang}}]{Han2023KaiC}%
  \BibitemOpen
  \bibfield  {author} {\bibinfo {author} {\bibfnamefont {X.}~\bibnamefont
  {Han}}, \bibinfo {author} {\bibfnamefont {D.}~\bibnamefont {Zhang}}, \bibinfo
  {author} {\bibfnamefont {L.}~\bibnamefont {Hong}}, \bibinfo {author}
  {\bibfnamefont {D.}~\bibnamefont {Yu}}, \bibinfo {author} {\bibfnamefont
  {Z.}~\bibnamefont {Wu}}, \bibinfo {author} {\bibfnamefont {T.}~\bibnamefont
  {Yang}}, \bibinfo {author} {\bibfnamefont {M.}~\bibnamefont {Rust}}, \bibinfo
  {author} {\bibfnamefont {Y.}~\bibnamefont {Tu}},\ and\ \bibinfo {author}
  {\bibfnamefont {Q.}~\bibnamefont {Ouyang}},\ }\bibfield  {title} {\bibinfo
  {title} {Determining subunit-subunit interaction from statistics of cryo-em
  images: observation of nearest-neighbor coupling in a circadian clock protein
  complex},\ }\href {https://doi.org/10.1038/s41467-023-41575-1} {\bibfield
  {journal} {\bibinfo  {journal} {Nature Communications}\ }\textbf {\bibinfo
  {volume} {14}},\ \bibinfo {pages} {5907} (\bibinfo {year}
  {2023})}\BibitemShut {NoStop}%
\bibitem [{\citenamefont {Fang}\ \emph {et~al.}(2022)\citenamefont {Fang},
  \citenamefont {Hon}, \citenamefont {Zhou},\ and\ \citenamefont
  {Lu}}]{fang_empirical_2022}%
  \BibitemOpen
  \bibfield  {author} {\bibinfo {author} {\bibfnamefont {R.}~\bibnamefont
  {Fang}}, \bibinfo {author} {\bibfnamefont {J.}~\bibnamefont {Hon}}, \bibinfo
  {author} {\bibfnamefont {M.}~\bibnamefont {Zhou}},\ and\ \bibinfo {author}
  {\bibfnamefont {Y.}~\bibnamefont {Lu}},\ }\bibfield  {title} {\bibinfo
  {title} {An empirical energy landscape reveals mechanism of proteasome in
  polypeptide translocation},\ }\href {https://doi.org/10.7554/eLife.71911}
  {\bibfield  {journal} {\bibinfo  {journal} {eLife}\ }\textbf {\bibinfo
  {volume} {11}},\ \bibinfo {pages} {e71911} (\bibinfo {year}
  {2022})}\BibitemShut {NoStop}%
\bibitem [{\citenamefont {Hohenberg}\ and\ \citenamefont
  {Halperin}(1977)}]{hohenberg_theory_1977}%
  \BibitemOpen
  \bibfield  {author} {\bibinfo {author} {\bibfnamefont {P.~C.}\ \bibnamefont
  {Hohenberg}}\ and\ \bibinfo {author} {\bibfnamefont {B.~I.}\ \bibnamefont
  {Halperin}},\ }\bibfield  {title} {\bibinfo {title} {Theory of dynamic
  critical phenomena},\ }\href {https://doi.org/10.1103/RevModPhys.49.435}
  {\bibfield  {journal} {\bibinfo  {journal} {Rev. Mod. Phys.}\ }\textbf
  {\bibinfo {volume} {49}},\ \bibinfo {pages} {435} (\bibinfo {year}
  {1977})}\BibitemShut {NoStop}%
\bibitem [{\citenamefont {Tu}(2023)}]{tu_renormalization_2023}%
  \BibitemOpen
  \bibfield  {author} {\bibinfo {author} {\bibfnamefont {Y.}~\bibnamefont
  {Tu}},\ }\bibfield  {title} {\bibinfo {title} {The renormalization group for
  non-equilibrium systems},\ }\href
  {https://doi.org/10.1038/s41567-023-02196-z} {\bibfield  {journal} {\bibinfo
  {journal} {Nat. Phys.}\ }\textbf {\bibinfo {volume} {19}},\ \bibinfo {pages}
  {1536} (\bibinfo {year} {2023})}\BibitemShut {NoStop}%
\end{thebibliography}%
\end{document}